\documentclass[showpacs, aps, superscriptaddress, twocolumn]{revtex4}
\usepackage{mathrsfs}
\usepackage{array}
\usepackage{amsmath}
\usepackage{amssymb}
\usepackage{graphicx}
\usepackage{color,xcolor}
\usepackage{booktabs}
\usepackage{amstext}
\DeclareMathOperator{\Tr}{Tr}
\usepackage{amsfonts}
\usepackage{float}
\usepackage{bm}
\usepackage{epstopdf}
\usepackage{color}

\begin{document}

\title{Probing topological states through the exact non-Markovian decoherence dynamics of a spin coupled to a spin bath in real-time domain}
\author{Chuan-Zhe Yao}
\affiliation{Department of Physics and Center for Quantum
information Science, National Cheng Kung University, Tainan 70101,
Taiwan}
\author{Wei-Min Zhang}
\email{wzhang@mail.ncku.edu.tw}
\affiliation{Department of Physics
and Center for Quantum information Science, National Cheng Kung
University, Tainan 70101, Taiwan}
\begin{abstract}
In this paper, we explore the decoherence dynamics of a probing spin coupled to a spin bath, where the spin bath is given 
by a controllable 1D transverse-field Ising chain. The 1D transverse-field Ising chain with free-ends boundary condition is 
equivalent to a modified Kitaev model with non-local Majorana bound states in its topological phase. We find that the non-Markovian 
decoherence dynamics of the probing spin can manifest the topological structure of the spin chain. By controlling the external 
magnetic field on the Ising chain, we find a close relationship between the topological phase transition and the non-Markovian 
dynamics in the real-time domain. We also investigate the corresponding quantum entanglement 
dynamics in this topological system.
\end{abstract}
\maketitle
\section{Introduction}
In condensed matter physics, the transverse-field Ising model (equivalently the Heisenberg-Ising chain) not only allows 
identification of quantum phase transitions \cite{Lieb61,1970}, but also has been experimentally realized through the 
CoNb$_{2}$O$_{6}$ compound \cite{realize1}, trapped ions \cite{trapion1,trapion2,trapion3,trapion4}, Mott insulator 
\cite{realize2}, and Rydberg atom \cite{realize3} etc., and therefore it has been widely investigated.  
On the other hand, recent experiments with polar molecules and ion chains provide a new direction for the dynamics of 
many-body systems in the real-time domain. In particular, the transverse-field Ising model has been revisited in the investigation of 
non-equilibrium physics, such as dynamical quantum phase transition, through the time evolution of observables in the transverse-field 
Ising chain under dynamical quench \cite{quench1,quench2,quench3,quench4}, the Loschmidt echo of a probing spin homogeneously 
coupled to a transverse-field Ising chain \cite{LE1,LE2,LE3}, and the decoherence dynamics of a transverse-field Ising 
chain coupled to a thermal bath \cite{thermalbath1,thermalbath2}. However, most of these investigations are mainly 
considered for Markov processes, while the dynamics of many open systems are often non-Markovian dominanted. In this paper, 
we shall investigate the real-time non-Markovian dynamics for a transverse-field Ising model in different quantum phases with different initial states, 
through its coupling to a probing spin.

As it is well known, by Jordan-€"Wigner transformation, the transverse-field Ising model can be mapped onto the Kitaev 
chain model \cite{Kitaev2001,MajoReview,IsingtoMajo,IsingtoKitaev}. In the fermionic representation, the well-known quantum phase 
transition of the model can be understood as a transition from the weak-pairing BCS regime to the strong-pairing Bose-Einstein 
condensate regime \cite{BEC-BCS,MajoIsing}. The phase diagram can be classified according to the topological 
order \cite{BEC-BCS,MajoIsing,WindIsing1,WindIsing2}. 
Moreover, the Kitaev model possesses Majorana zero modes (Majorana bound states) non-locally separated at the two ends 
of the open chain in the topologically non-trivial phase. The dynamical behavior of quantum phase transition in the model must 
relate topologically to the non-local property of the Majorana zero modes. However, the previous studies mainly consider the 
transverse-field Ising model with periodic boundary condition where the Majorana zero modes cannot be manifested. Meanwhile, 
even though the solution of the transverse-field Ising model with free-end boundary condition is exactly solvable, its eigenenergies and eigenfunctions are 
determined by a transcendental equation which has not been analytically solved so far  \cite{Lieb61,1970,thermalbath2}.

In this paper, we will modify the transverse-field Ising model such that the local magnetic field does not apply to the last spin 
of the Ising chain. We find that such modified model can be solved analytically with the free-end boundary condition for the 
eigenenergies and eigenfunctions. Moreover, 
we derive the exact master equation of a probing spin coupled to the transverse-field Ising model \cite{PRB2008,NewJ2010,PRA2010,Opt2010,PRL2012,Ann2012,PRB2015,Pei2018,PRB2018}. 
Through the investigation of  non-Markovian dynamics of the probing spin coupled to this modified transverse-filed Ising chain, 
one can probe experimentally,  for example with Ramsey interferometry,  how the non-trivial topological properties of the transverse-filed Ising model 
 can be manifested in the real-time domain. A great number of papers have been devoted to the study entanglement close 
to topological phase transition, and there have been indications that entanglement is enhanced near the quantum critical point 
\cite{Nature2002,PRA2002,PRL2003,Cardy2004}. We also numerically explore the non-Markovian dynamics of the entanglement entropy which shows 
a diagnostic tool for the study of topological phase transitions.

The rest of the paper is organized as follows. In Sec.~\ref{sec2}, we introduce our modified transverse-field Ising model and study 
its topological characterization. In Sec.~\ref{sec3}, we derive the exact master equation of a probing spin couple to the modified 
transverse-field Ising model using the path integral approach in the coherent state representation \cite{PRB2008}. In Sec.~\ref{sec4}, 
we analyze in detail the non-Markovian decoherence dynamics of the probing spin coupled to the modified transverse-field Ising 
model by investigating two-time spin-spin correlation functions in different phases of the spin chain, different spin-spin chain 
coupling, different spin-flip energy, and different initial states. The effects of topologically non-local property on the non-Markovian 
dynamics are also clarified under different conditions. In Sec.~\ref{sec5}, we solve the dynamics of the entanglement entropy 
through the exact master equation in Sec.~\ref{sec3}. We also find the close relationships between the dynamical entanglement 
entropy and the topological phase transition in the real-time domain. The relation between the entanglement entropy and the 
two-time spin-spin correlation obtained in Sec.~\ref{sec4} is also presented. Finally, a conclusion is given in Sec.~\ref{sec6}. 
The detailed derivations of the formulas are presented in the appendices.

\section{The modified transverse-field Ising model and its topological structure}
\label{sec2}
\begin{figure}[t]
\centerline{\scalebox{0.35}{\includegraphics{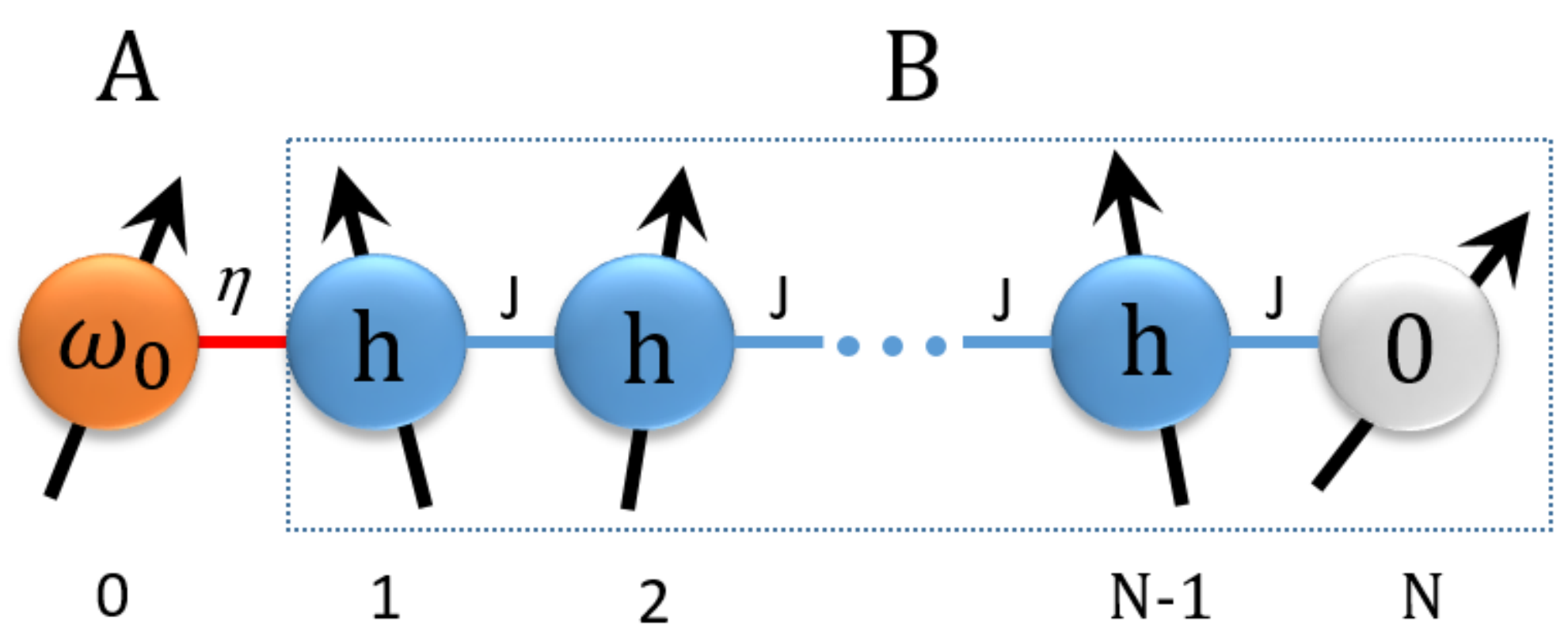}}}
\caption{(Colour online) A schematic diagram of the probing spin coupled to the modified transverse-field Ising chain with setting 
the parameters of the coupling constant and the external magnetic field as in Eq.~(\ref{H0}).}\label{fig1}
\end{figure}

To probe topological structure and dynamical phase transition through non-Markovian decoherence dynamics 
and entanglement entropy in the real-time domain, we couple a probing spin to a modified 1D transverse-field Ising chain as shown in Fig.~\ref{fig1}. 
The Hamiltonian of the system is
\begin{align}
H&=H_{A}+H_{I}+H_{B}\notag\\
&= - \omega_{0}\sigma_{0}^{z}-\eta\sigma_{0}^{x}\sigma_{1}^{x}-\sum\limits_{j=1}^{N-1}J_{j}\sigma_{j}^{x}\sigma_{j+1}^{x}-\sum\limits_{j=1}^{N}h_{j}\sigma_{j}^{z} ,
\label{H0}
\end{align}
where the first term is the Hamiltonian of the probing spin $\sigma_{0}$, the second term is the coupling between the probing spin $\sigma_{0}$ 
and the first spin $\sigma_{1}$ in the modified transverse-field Ising chain. The last two terms are the Hamiltonian of the modified 
1D transverse-field Ising model, which is an N-site spin chain with the nearest coupling $J_{j}$ and the local external magnetic field $h_{j}$
with $\sigma^{x,y,z}_j$ being the Pauli matrices.
Different from the previous works on the conventional transverse-field Ising chain with a period boundary condition, here we set the 
transverse-field Ising chain to have a free-end boundary condition so that its topological features can be presented on the edges. 
However, as shown as early by Lieb {\it et al.} \cite{Lieb61} and later by Pfeuty \cite{1970}, although the transverse-field Ising model with free ends is exactly 
solvable, its  eigenenergies and eigenfunctions are determined by a transcendental equation which cannot be solved analytically.
Hence, we modify the model by setting $h_{N}=0$, namely, the local magnetic fields do not apply to the spin $\sigma_{N}$ at last site. This modification makes 
the free-boundary transverse-field Ising chain analytically solvable, and meantime the topological structure of the system can still be maintained 
as we will show later. The spin-flip energy $\omega_{0}$ of the probing spin $\sigma_{0}$ and the coupling energy $\eta$ between 
$\sigma_{0}$ and $\sigma_{1}$ are controllable. For simplicity, we also set $J_{i}=J$, $h_{i}=h$ ($i=1,\cdots,N-1$).

By applying the Jordan-Wigner transformation
\begin{subequations}
\begin{align}
\sigma_{j}^{+}&=(\sigma_{j}^{x}+i\sigma_{j}^{y})/2=c_{j}^{\dagger}\prod\limits_{m<j}e^{-i\pi c_{m}^{\dagger}c_{m}}\\
\sigma_{j}^{-}&=(\sigma_{j}^{x}-i\sigma_{j}^{y})/2=c_{j}\prod\limits_{m<j}e^{i\pi c_{m}^{\dagger}c_{m}}\\
\sigma_{j}^{z}&=2c_{j}^{\dagger}c_{j}-1,
\end{align}
\end{subequations}
the total system can be transformed into a fermionic system:
\begin{align}
H&= - \bold\omega_{0}(2a^{\dagger}a -1) -\eta(a^{\dagger}-a)(c^{\dagger}_{1}+c_{1})\notag\\
&-\sum\limits_{j=1}^{N-1}(Jc^{\dagger}_{j}c^{\dagger}_{j+1}+Jc^{\dagger}_{j}c_{j+1}+ hc_{j}^{\dagger}c_{j}+H.c.). \label{km}
\end{align}
Here we have ignored a constant term in the above Hamiltonian.
As it is shown, after the Jordan-Wigner transformation, the transverse-field Ising chain is reduced to the Kitaev chain with the same hoping 
and pairing strengths \cite{Lieb61,1970,Kitaev2001,MajoIsing}, except that the on-site chemical potential of the last site, $j=N$,
vanishes as an effect of the modification of the model, see Eq.~(\ref{km}). 
We take further a Bogoliubov transformation to the spin chain
\begin{subequations}
\label{B0}
\begin{align}
b_{k}&=\sum\limits_{i=1}^{N}(u_{ki}c_{i}+v_{ki}c_{i}^{\dagger})\\
b_{k}^{\dagger}&=\sum\limits_{i=1}^{N}(v_{ki}^{*}c_{i}+u_{ki}^{*}c_{i}^{\dagger}),
\end{align}
\end{subequations}
such that $H_{B}$ is diagonalized,
\begin{align}
H_{B}=\sum\limits_{k}\epsilon_{k}(b_{k}^{\dagger}b_{k}-b_{k}b_{k}^{\dagger})
\label{B1}
\end{align}
where $b_{k}$ and $b^{\dagger}_{k}$ are creation and annihilation operators of Bogoliubov quasi-particles (bogoliubons) with the spectrum
\begin{align}
\epsilon_{k}=\left\{\begin{array}{ll}
J\sqrt{1+\lambda^{2}-2\lambda\cos\frac{k\pi}{N}} , & k =1, 2, \cdots, N-1 \\~\\
0 , & k=k_0
\end{array} \right. ,  
\label{e}
\end{align}
where $\lambda=h/J$, and the zero energy mode $k_0$ is determined by $1+\lambda^{2}-2\lambda\cos\frac{k_0\pi}{N}=0$.
  The corresponding wavefunctions for the non-zero energy bogoliubons can be analytically solved
\begin{subequations}
\label{kmode}
\begin{align}
u_{kj}&={\cal N}_{k}\bigg\{\frac{-J}{\epsilon_{k}}\sin\bigg[\dfrac{(j-1)k\pi}{N}\bigg]+\bigg(1-\dfrac{J\lambda}{\epsilon_{k}}\bigg)\sin \dfrac{jk\pi}{N}\bigg\} , \\
v_{kj}&={\cal N}_{k}\bigg\{\frac{-J}{\epsilon_{k}}\sin\bigg[\dfrac{(j-1)k\pi}{N}\bigg]-\bigg(1+\dfrac{J\lambda}{\epsilon_{k}}\bigg)\sin \dfrac{jk\pi}{N}\bigg\},
\end{align}
\end{subequations}
and that of the zero-energy bogoliubon is
\begin{subequations}
\label{zero}
\begin{align}
u&_{k_0j}=\left\{\begin{array}{l}
{\cal N}_{k_0}(-\lambda)^{j-1}\hfill j<N\\~\\
{\cal N}_{k_0}(-\lambda)^{N-1}+1/2\quad\hfill j=N
\end{array} \right.\\
v&_{k_0j}=\left\{\begin{array}{l}
{\cal N}_{k_0}(-\lambda)^{j-1}\hfill j<N\\ ~\\
{\cal N}_{k_0}(-\lambda)^{N-1}-1/2\quad\hfill j=N
\end{array} \right. ,
\end{align}
\end{subequations}
where ${\cal N}_{k},{\cal N}_{k_0}$ are the normalization constants which are given by
\begin{subequations}
\begin{align}
{\cal N}_{k}&=\left\{N\left[1+2\dfrac{J^{2}}{\epsilon^{2}_{k}}\Big(1+\cos \dfrac{k\pi}{N}\Big)\right]\right\}^{-1/2}\\
{\cal N}_{k_0}&=\dfrac{1}{2}\left(\dfrac{1-\lambda^{2N}}{1-\lambda^{2}}\right)^{-1/2} .
\end{align}
\end{subequations}
The detailed derivation is given in Appendix A. 
Thus, the Hamiltonian of Eq.~(\ref{H0}) can be expressed as
\begin{align}
H=-&\bold\omega_{0}(2a^{\dagger}a-1)-\sum\limits_{k}V_{k}(a^{\dagger}-a)(b^{\dagger}_{k}+b_{k})\notag\\
+&\sum\limits_{k}\epsilon_{k}(2b_{k}^{\dagger}b_{k}-1) ,
\label{H}
\end{align}
with
\begin{align}
V_{k}=\left\{\begin{array}{ll}
\dfrac{-2\eta\lambda\sin\frac{k\pi}{N}}{\sqrt{N(1+\lambda^{2}-2\lambda\cos\frac{k\pi}{N})}} , &  k =1, 2, \cdots, N-1 \\ ~\\
\sqrt{2}\eta\bigg(\sum\limits_{j=0}^{N-1}\lambda^{2j}\bigg)^{-1/2} , & k=k_0\\
\end{array} \right. .
\label{z}
\end{align}

\begin{figure}[t]
\includegraphics[scale=0.3]{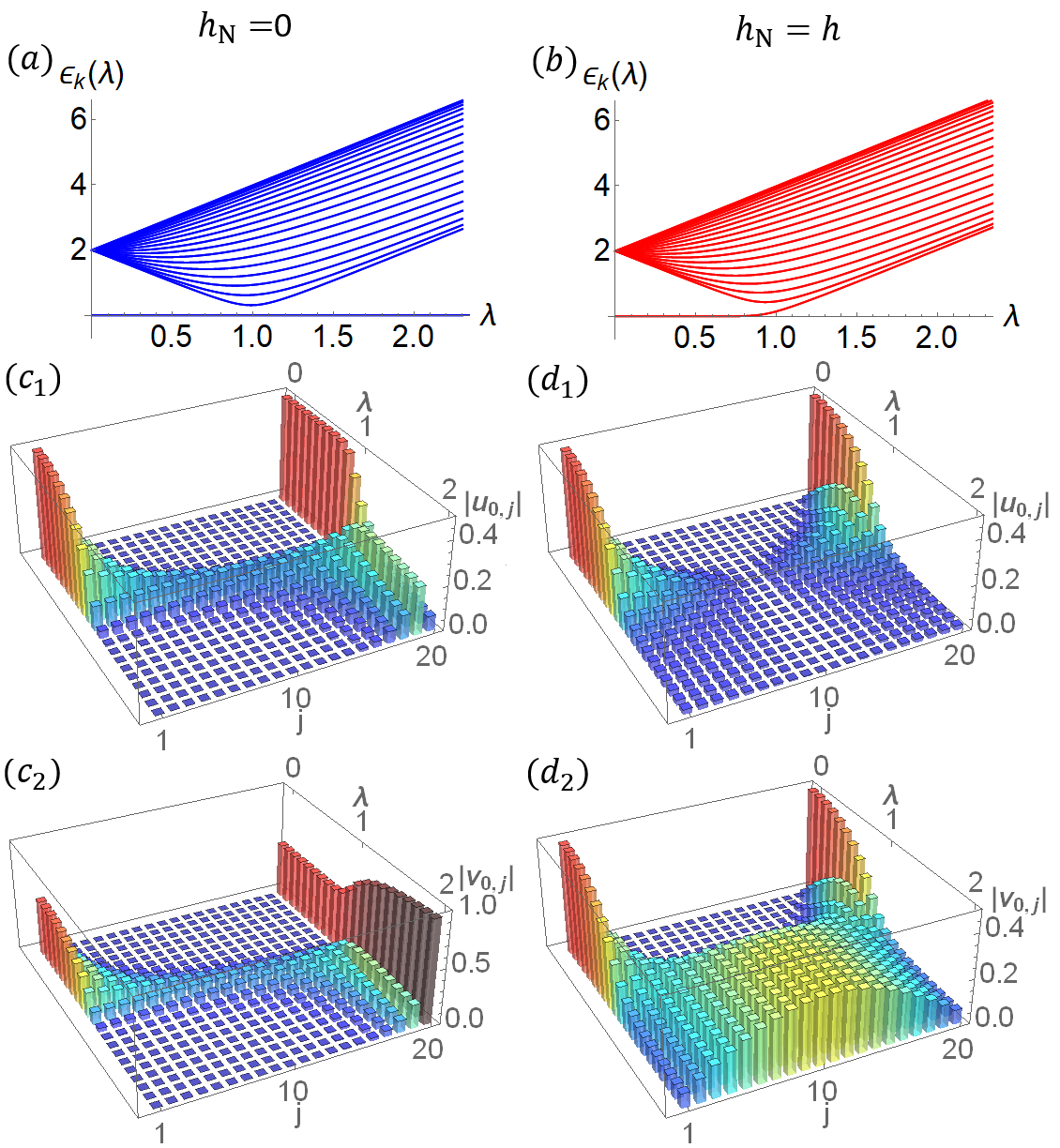}
\caption{(Colour online) The spectrum for (a) the modified transverse-field Ising chain ($h_{N}=0$) and (b) the ordinary 
transverse-field Ising chain ($h_{N}=h$) with $N=20$. The wavefunction distribution of the ground state, $|u_{0,j}|$ (top) 
and $|v_{0,j}|$ (bottom) for (c) $h_{N}=0$ and (d) $h_{N}=h$ with varying from the topologically non-trivial phase $\lambda=0$ 
to the topologically trivial phase $\lambda=2$.} \label{fig2}
\end{figure}

The above analytical solution, Eq.~(\ref{e}-\ref{zero}), is in fact a consequence of the modification with $h_{N}=0$. 
In the case of $h_{N}=h$, no such analytical solution has been found in the literature (see Appendix A). Meanwhile, the modified 
transverse-field Ising chain has some different characters from the Kitaev chain. The difference is manifested first 
in the spectra of the model with and without setting $h_{N}=0$, as shown in Fig.~\ref{fig2}(a) and Fig.~\ref{fig2}(b), 
respectively. The resulting excited state spectra are similar for the two cases, but their ground state behavior is very different. 
The modified model always has zero-energy states, independent of the value of $\lambda$. While the zero-energy states 
only exist in the region of $\lambda<1$ for large $N$ in the ordinary transverse-field Ising chain (or the equivalent Kitaev chain).

Secondly and more importantly, the wavefunction distribution of the ground states in the two cases are significantly different, except for  
$\lambda=0$ (no transverse field), as shown in Fig.~\ref{fig2}$(c_1)$, $(c_2$), ($d_1$) and ($d_2$). Note that the zero-energy bogoliubon state ($\epsilon_k=0$) is two-fold 
degenerate, with the particle number $b^{\dagger}_{0}b_{0}=0$ and $1$, respectively. These two states can be 
described by the left and right Majorana operators $\gamma_{L}=-i(b_{0}-b_{0}^{\dagger}),\gamma_{R}=b_{0}+b_{0}^{\dagger}$ 
\cite{Kitaev2001}. Figure \ref{fig2}$(c_1)$-$(c_2)$ demonstrates the non-local separation 
of the two Majorana zero modes in the topologically non-trivial phase ($\lambda<1$), distributed asymmetrically in the 
two sides of the spin chain. In particular, if $\lambda=0$, 
these two Majorana zero modes locate perfectly at end of the two sides of the spin chain, just the same as that in Kitaev 
model \cite{Kitaev2001}. As $\lambda$ gets increase, only the left Majorana zero mode wavefunction spreads into other sites, 
while the right Majorana zero mode remains unchanged due to the setting $h_{N}=0$, as shown in Fig.~\ref{fig2}(c). 
At the critical point $\lambda_{c}=1$, the left Majorana zero mode wavefunction is uniformly distribute over the all sites of the chain but
the right Majorana zero mode still remains unchanged. 
However, for the Kitaev chain, the wavefunction of the left and right Majorana zero modes are symmetrically distributed 
over the chain for $\lambda < 1$, see Fig.~\ref{fig2}(d). By comparing Fig.~\ref{fig2}(c) with Fig.~\ref{fig2}(d), 
we find that the wavefunction distribution of the left Majorana zero mode are the same for $\lambda < 1$ in both cases,
but the wavefunction distribution of the right Majorana zero mode are very different. 
When $\lambda>1$, for the modified transverse-field Ising chain, the wavefunction of the left Majorana zero mode 
distributes more on the right-hand side (r.h.s.) than the left-hand side (l.h.s.). With continuously increasing $\lambda$,  the left  
Majorana zero mode $\gamma_{L}$ eventually condenses with the right Majorana zero mode $\gamma_R$ to the last 
site $N$ so that although the Majorana zero-modes still exist but no longer have the topologically non-local property. This solution ($\lambda>1$) 
of the modified Ising chain is very different from the Kitaev model in which there exists no longer zero-energy state 
for $\lambda>1$, as shown in Fig.~\ref{fig2}.  

We summarize the above topological properties of the modified transverse-field Ising chain as follows: the condition $h_N=0$ makes the right 
Majorana zero mode always localize at the end of the r.h.s. of spin chain, independent of the value of $\lambda$. 
Meanwhile, the wavefunction distribution of the left Majorana zero mode changes and moves from the l.h.s.~to the r.h.s.~of 
the spin chain when the parameter $\lambda$ changes from $\lambda < 1$ to $\lambda > 1$, which results in a topological 
phase transition at the critical point $\lambda_c=1$.  This topological phase transition remains unchanged even in the limit $N \rightarrow \infty$, 
because it is the local and non-local topological properties of the Majorana zero mode wavefunctions associated with the ends  
of the spin chain rather than its length. In fact, this topological feature becomes more significant for the larger $N$, where the 
non-locality of Majorana zero modes is manifested clearer \cite{Kitaev2001}.
On the other hand, practically $h_N$ may not be ideally zero, i.e., it may have some small but non-zero 
local transverse field $h_N$ in experiments. However, a very small $h_N$ only causes a very small tail to the wavefunction distribution of the 
right Majorana zero mode over a couple of sites from the right end of the spin chain. This small wavefunction tail does not change the above 
topological feature, as an evidence of topological protection from local perturbation \cite{Kitaev2001}.
 
The above topological properties of the zero-energy states can be understood more comprehensively through 
calculating the winding number, which is used to identify topological phases of matter \cite{MajoIsing,WindIsing2,Wind1,Wind2}. To this end, we rewrite the Hamiltonian of the spin chain in Eq.~(\ref{H}) in the pseudo spin representation:
\begin{align}
H_{B}=\sum\limits_{k}
\begin{pmatrix}
c_{k}^{\dagger}&c_{k}
\end{pmatrix}
P^{-1}
\begin{pmatrix}z(k)&x(k)\\
x(k)&-z(k)\end{pmatrix}
P\begin{pmatrix}c_{k}\\c_{k}^{\dagger}\end{pmatrix},
\end{align}
where
\begin{align}
P=\begin{pmatrix}z(k)-\sqrt{z^{2}(k)+x^{2}(k)}&x(k)\\
z(k)+\sqrt{z^{2}(k)+x^{2}(k)}&x(k)\end{pmatrix}
\end{align}
and
\begin{subequations}
\begin{align}
x(k)&=J\sin(k\pi/N)\\
z(k)&=J\cos(k\pi/N)-h+(h-j)\delta_{k,k_0}.
\end{align}
\end{subequations}
The winding number is defined as the line integral along a close curve on the $z-x$ plane
\begin{align}
W=\frac{1}{2\pi}\int_{c}\frac{1}{x^{2}+z^{2}}(zdx-xdz),
\end{align}
as the total number of times that the curve travels counterclockwise around the origin. Explicit calculation shows that 
the winding number $W=1$ for $\lambda<1$, which means that the spin chain is in the topologically non-trivial phase, 
while it is in the topologically trivial phase with $W=0$ for $\lambda>1$, although there is still zero-energy ground state. 
This demonstrates a topological phase transition in the modified transverse-field Ising chain associated with the topological 
non-local feature, namely, a transition from the topologically nontrivial phase to the topologically trivial phase occurs 
when $\lambda$ passes through $\lambda_{c}=1$.

In conclusion, the modified transverse-field Ising chain exhibits a similar topological phase transition as the ordinary 
transverse-field Ising chain, but the ground state energy and its wavefunctions behave so different 
in the two models. It also demonstrates explicitly that the topology of the system is determined by the detailed 
non-local properties of the zero-mode wavefunctions, rather than the system spectra. Because the modified transverse-field Ising model can 
be analytically solved explicitly for both the eigenenergies and eigenwavefunctions, we can also use 
it to study the exact decoherence dynamics of the system through its coupling to a probing spin, which is fully encapsulated 
in the spectral density $J(\omega)\equiv 2\pi \sum_k |V_k|^2 \delta(\omega_k-\omega)= 2\pi \rho(\omega) |V(\omega|^2$. Here
$\rho(\omega)$ is the density of states of the spin chain that can be determined from Eq.~(\ref{e}), and 
$V(\omega)$ is the coupling amplitude of the probing spin coupled to the spin chain that involves explicitly all the 
eigenwavefunction distributions of the spin chain as given by Eq.~(\ref{z}).  This indicates that
the topological properties of the spin chain can be experimentally observed from the decoherence dynamics of the probing spin, as 
we shall show in the next sections.

\section{The exact master equation}
\label{sec3} 
\begin{figure}[H]
\centerline{\scalebox{0.28}{\includegraphics{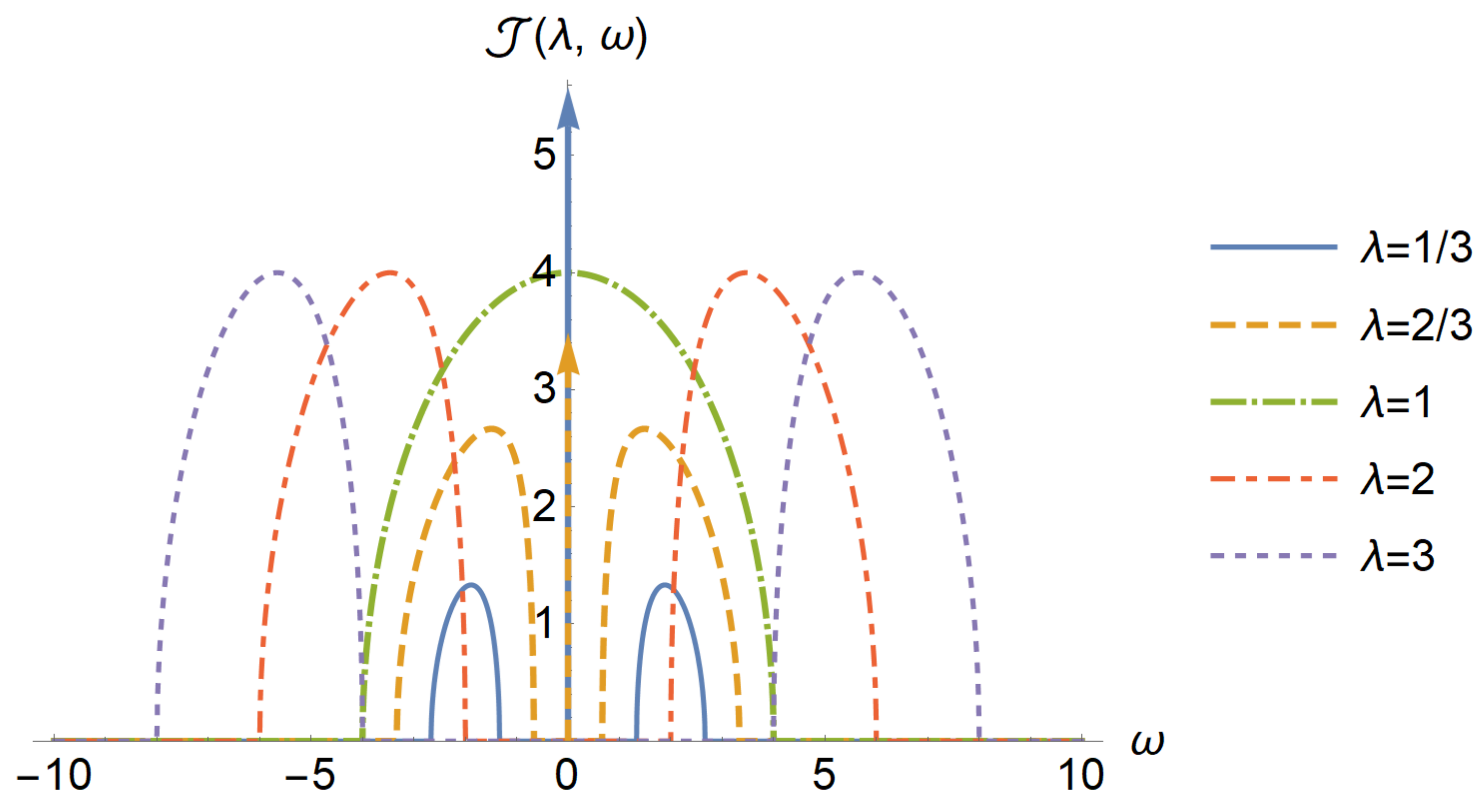}}}
\caption{(Colour online) The effective spectral density $\mathcal{J}(\lambda,\omega)$ from the topologically 
non-trivial phase ($\lambda<1$) to the topologically trivial phase ($\lambda>1$).}\label{fig3}
\end{figure}

The topological properties and topological phase transition of the modified Ising chain can be explored through 
the non-Markovian decoherence dynamics of the probing spin, which is described by the time evolution 
of the reduced density matrix of the probing spin. The reduced density matrix is obtained from the total density matrix of the probing spin 
and the spin chain by tracing out all possible states of the spin chain
\begin{align}
\rho_{A}(t)=\Tr_{B}[U(t,t_{0})\rho_{tot}(t_{0})U^{\dagger}(t,t_{0})],
\end{align}
where $U(t,t_{0})=\exp[-iH(t-t_{0})]$ is the time evolution operator of the total system. Initially we assume that 
the two subsystem (spin $\sigma_{0}$ and the spin chain) are decoupled \cite{Feynman1963,Leggett}, 
that is, $\rho_{tot}(t_{0})=\rho_{A}(t_{0})\otimes\rho_{B}(t_{0})$. Then in the fermionic coherent state representation
\begin{align}
\langle\xi_{f}&|\rho_{A}(t)|\xi_{f}^{\prime}\rangle\notag\\
=&\int d\mu(\xi_{0})d\mu(\xi_{0}^{\prime})\langle\xi_{0}|\rho_{A}(t_{0})|\xi_{0}^{\prime}\rangle\mathcal{K}(\xi_{f}^{*},\xi_{f}^{\prime},t|\xi_{0},\xi_{0}^{\prime*},t_{0}),
\label{rho0}
\end{align}
where $\xi_{0}$, $\xi_{0}^{\prime*}$, $\xi_{f}^{\prime}$, $\xi_{f}^{*}$ are the eigenvalues of the fermionic 
coherent states and are Grassmann numbers. The propagator $\mathcal{K}(\xi_{f}^{*},\xi_{f}^{\prime},t|\xi_{0},\xi_{0}^{\prime*},t_{0})$ is 
determined by the action of the probing spin $\sigma_{0}$ and the influence functional arose from the spin chain \cite{Feynman1963}, 
the later is obtained by integrating out all the degree of freedom of the 
spin chain \cite{PRB2008,NewJ2010,PRA2010,Opt2010,PRL2012,Ann2012,PRB2015,Pei2018,PRB2018}. 
With a tedious derivation (see appendix B), the exact master equation 
for the probing spin coupling to the transverse-field Ising chain is obtained
\begin{align}
\dot{\rho}_{A}(t)
=-&i[\epsilon(t,t_{0})a^{\dagger}a,\rho_{A}(t)]\notag\\
+&\gamma(t,t_{0})[2a\rho_{A}(t)a^{\dagger}-a^{\dagger}a\rho_{A}(t)-\rho_{A}(t)a^{\dagger}a]\notag\\
+&\tilde{\gamma}(t,t_{0})[a^{\dagger}\rho_{A}(t)a-a\rho_{A}(t)a^{\dagger} +a^{\dagger}a\rho_{A}(t) \notag\\
&~~~~~~~~~~ -\rho_{A}(t)aa^{\dagger} ] \notag\\
+&\Lambda(t,t_{0})a^{\dagger}\rho_{A}(t)a^{\dagger}+\Lambda^{*}(t,t_{0})a\rho_{A}(t)a,
\label{master eq}
\end{align}
where all the time-dependent coefficients are determined by the generalized non-equilibrium Green functions incorporating the pairing dynamics as follows,
\begin{subequations}
\begin{align}
\epsilon(t,t_{0})=&\dfrac{i}{2}[\dot{\boldsymbol{U}}(t,t_{0})\boldsymbol{U}^{-1}(t,t_{0})-H.c]_{11}, \\
\gamma(t,t_{0})=&-\dfrac{1}{2}[\dot{\boldsymbol{U}}(t,t_{0})\boldsymbol{U}^{-1}(t,t_{0})+H.c]_{11},\\
\tilde{\gamma}(t,t_{0})=&\dot{\boldsymbol{V}}_{11}(t,t)-[\dot{\boldsymbol{U}}(t,t_{0})\boldsymbol{U}^{-1}(t,t_{0})\boldsymbol{V}(t,t)+H.c.]_{11},\\
\Lambda(t,t_{0})=&-[\dot{\boldsymbol{U}}(t,t_{0})\boldsymbol{U}^{-1}(t,t_{0})]_{12} .
\end{align}
\end{subequations}
The Green functions $\boldsymbol{U}(t,t_{0})$ and $\boldsymbol{V}(t,t)$ are $2\times2$ matrix and satisfy 
the integro-differential equations \cite{PRB2008,NewJ2010,PRA2010,Opt2010,PRL2012,Ann2012,PRB2015,Pei2018,PRB2018}
\begin{subequations}
\begin{align}
\dfrac{d}{dt}\boldsymbol{U}(t,t_{0})&-2i\bold\omega_{0}\begin{pmatrix}
1&0\\
0&-1
\end{pmatrix}\boldsymbol{U}(t,t_{0})\notag\\
&+\int_{t_{0}}^{t}\boldsymbol{G}(t,\tau)\boldsymbol{U}(\tau,t_{0})d\tau=0\label{U}\\
\boldsymbol{V}(t,\tau)=&
\int_{t_{0}}^{\tau}d\tau_{1}\int_{t_{0}}^{t}d\tau_{2}\boldsymbol{U}(\tau,\tau_{1})
\tilde{\boldsymbol{G}}(\tau_{1},\tau_{2})\boldsymbol{U}^{\dagger}(t,\tau_{2})\label{V}
\end{align}
\end{subequations}
with the initial condition $\boldsymbol{U}(t_{0},t_{0})=\boldsymbol{I}$. The integral memory kernels
\begin{subequations}
\begin{align}
\boldsymbol{G}(t,t_{0})=&2\operatorname{Re}[g(t,t_{0})]\begin{pmatrix}
1&-1\\
-1&1
\end{pmatrix}\label{GU}\\
\tilde{\boldsymbol{G}}(t,t_{0})=&\{g(t,t_{0})-2\operatorname{Im}[g_{\beta}(t,t_{0})]\}\begin{pmatrix}
1&-1\\
-1&1
\end{pmatrix},
\end{align}
\end{subequations}
where
\begin{subequations}
\begin{align}
g(t,t_{0})=&\int\frac{d\omega}{2\pi}J(\omega)e^{-i\omega(t-t_{0})},\\
g_{\beta}(t,t_{0})=&\int\frac{d\omega}{2\pi}J(\omega)f(\omega)e^{-i\omega(t-t_{0})},
\label{g}
\end{align}
\end{subequations}
and $f(\omega)=[e^{\beta(\epsilon_{k}-\mu)}+1]^{-1}$ is the Fermi-Dirac distribution of the spin chain 
at initial time $t_{0}$. The spectral density of the system,
\begin{align}
J(\omega)\equiv 2\pi\sum_{k}|V_{k}|^{2}\delta(\omega-\epsilon_{k})
=2\pi \rho(\omega)|V(\omega)|^2,
\end{align} 
where $V_{k}$ and $\epsilon_{k}$ are given by Eq.~(\ref{z}) and Eq.~(\ref{e}), respectively. The explicit form is given as follows,
\begin{align}
J(\omega)=&\frac{\eta^{2}}{\omega}\sqrt{-[\omega^{2}/4-J^{2}(1-\lambda)^{2}][\omega^{2}/4-J^{2}(1+\lambda)^{2}]}\notag\\
&+\left\{\begin{array}{l}
\pi\eta^{2}(1-\lambda^{2})\delta_{\omega,0}\quad\lambda<1\\~\\
0\hspace*{25.5mm}\lambda\geq1
\end{array} \right. ,
\label{spectral}
\end{align}
in which the last term is contributed from the non-local Majorana zero modes. Note that when $\lambda\geq1$, the zero modes 
have no contribution to the spectral density because the zero modes move to the right-hand side of the spin 
chain and therefore decouple from the probing spin. From Eq.~(\ref{U}) and (\ref{GU}), it shows that the 
memory kernel is determined by the effective spectral density $\mathcal{J}(\lambda, \omega)=2\operatorname{Re}[J(\omega)]$ 
plotted in Fig~\ref{fig3}. Notice that except for the case of $\lambda=1$, there is a gap in the middle of the effective 
spectral density, which will induce localized bound states and prevent decoherence \cite{PRL2012}, 
as we will discuss in detail in the next section.

\section{The exact non-Markovian dynamics}
\label{sec4}
\subsection{The analytical solution of the retarded and correlation Green functions}
By coupling the probing spin $\sigma_{0}$ with the spin chain (see Fig.~\ref{fig1}), we find that the dynamics 
of the probing spin manifest the topological properties of the spin chain. As one has seen, the renormalized 
Hamiltonian of the probing spin and the dissipation and fluctuation coefficients in its exact master equation 
Eq.~(\ref{master eq}) are all determined by the Green functions $\boldsymbol{U}(t,t_{0})$ and $\boldsymbol{V}(t,t)$. 
The solutions of these two Green functions fully depend on the density of states of the spin chain as well as 
the coupling between the probing spin and the spin chain through the spectral density of Eq.~(\ref{spectral}). 
Their physical consequences can be seen more clearly in the Heisenberg picture. After the Jordan-Wigner transformation, 
the dynamics of the probing spin $\sigma_{0}$ is described by the corresponding fermion operators $a(t)$ and $a^{\dagger}(t)$. 
Their Heisenberg equation of motions, after eliminating the degrees of freedom of the spin chain, lead to
\begin{align}
\frac{d}{dt}a(t)-&2i\omega_{0} a(t)
-2\int_{t_{0}}^{t}\operatorname{Re}[g(t,t_{0})][a^{\dagger}(\tau)-a(\tau)]d\tau\notag\\
=&\sum\limits_{k}iV_{k}[e^{-2i\epsilon_{k}t}b_{k}(t_{0})+e^{2i\epsilon_{k}t}b^{\dagger}_{k}(t_{0})]
\label{EOMa}
\end{align}
which is the generalized quantum Langevin equation \cite{PRB2015}, where the third term is a damping 
and the right-hand side of the equation is the noise force. Due to the linearity of Eq.~(\ref{EOMa}), its general solution has the form as
\begin{align}
\left(\begin{array}{c}
a(t)\\
a^{\dagger}(t)
\end{array}\right)=\boldsymbol{U}&(t,t_{0})\left(\begin{array}{c}
a(t_{0})\\
a^{\dagger}(t_{0})
\end{array}\right)\notag\\
+&\sum\limits_{k}\boldsymbol{F}_{k}(t,t_{0})\left(\begin{array}{c}
b_{k}(t_{0})\\
b_{k}^{\dagger}(t_{0})
\end{array}\right) ,
\label{EOM}
\end{align}
where $a(t_{0})$, $a^{\dagger}(t_{0})$, $b_{k}(t_{0})$ and $b_{k}^{\dagger}(t_{0})$ are the initial annihilation 
and creation operators of the probing spin $\sigma_{0}$ and the spin chain, respectively.

From Eq.~(\ref{EOM}), it can easily be shown that
\begin{align}
\boldsymbol{U}(t,t_{0})=\left( \begin{array}{cc}
\langle\{a(t),a^{\dagger}(t_{0})\}\rangle & \langle\{a(t),a(t_{0})\}\rangle\\
\langle\{a^{\dagger}(t),a^{\dagger}(t_{0})\}\rangle & \langle\{a^{\dagger}(t),a(t_{0})\}\rangle
\end{array}\right)
\end{align}
which is indeed an extension of the usual retarded Green function incorporating with pairings. The equation of 
motion of $\boldsymbol{U}(t,t_{0})$ is given by the integro-differential Eq.~(\ref{U}), which can also be easily 
justified by substituting Eq.~(\ref{EOM}) into the Heisenberg equation of motion (\ref{EOMa}). As shown in 
our previous work \cite{PRL2012}, the modified Laplace transform $\tilde{\boldsymbol{U}}(s)
=\int_{t_{0}}^{\infty}\boldsymbol{U}(t,t_{0})e^{is(t-t_{0})}$ of Eq.~(\ref{U}) is
\begin{align}
\tilde{\boldsymbol{U}}(s)=i\left( \begin{array}{cc}
s+2\omega_{0}-\Sigma(s) & \Sigma(s)\\
\Sigma(s) & s-2\omega_{0}-\Sigma(s)
\end{array}\right)^{-1},
\label{laplace U}
\end{align}
where the self-energy correction $\Sigma(s)$ is the Laplace transform of the integral kernel in Eq.~(\ref{U})
\begin{align}
\Sigma(s)=\int\dfrac{d\omega}{2\pi}\dfrac{\mathcal{J}(\lambda,\omega)}{s-\omega}\xrightarrow{s=\omega\pm i0^{+}}\Delta(\lambda,\omega)\mp\dfrac{i}{2}\mathcal{J}(\lambda,\omega),
\label{sigma}
\end{align}
and $\Delta(\lambda,\omega)=\mathcal{P}\int\frac{d\omega}{2\pi}\frac{\mathcal{J}(\lambda,\omega)}{s-\omega}$ is the principal 
value of the integral. Applying the inverse transformation to Eq.~(\ref{laplace U}), we can analytically solve 
$\boldsymbol{U}(t,t_{0})$, which consists of a summation of dissipationless oscillations arose from localized modes (localized bound 
states) determined by the real part of the self-energy correction) to the probing spin, plus nonexponential decays 
induced by the discontinuity of the imaginary part of the self-energy correction cross the real axes in the complex plane \cite{PRL2012}
\begin{align}
\boldsymbol{U}(t,t_{0})=\sum\limits_{s_{p}}&\left( \begin{array}{cc}
X(s_{p}) & Y(s_{p})\\
Y(s_{p}) & X(-s_{p})
\end{array}\right)e^{-is_{p}(t-t_{0})}\notag\\
+\int_{-\infty}^{\infty}&\dfrac{ds}{2\pi}
\dfrac{\mathcal{J}(s)e^{-is(t-t_{0})}}{[4\omega_{0}^{2}+(2\Delta(s)-s)s]^{2}+s^{2}(\mathcal{J}(s))^{2}}\notag\\[1mm]
&\times\left( \begin{array}{cc}
(s-2\omega_{0})^{2} & 4\omega_{0}^{2}-s^{2}\\
4\omega_{0}^{2}-s^{2} & (s+2\omega_{0})^{2}
\end{array}\right),
\label{exactU}
\end{align}
where $\{s_{p}\}$ is the set of the poles for the determinant of $\tilde{\boldsymbol{U}}(s)$ located at the real axis, i.e. $s-2\omega_{0}-\Delta(s_{p})=0$, and
\begin{subequations}
\begin{align}
X&(s)=\dfrac{[s-2\omega_{0}-\Delta(s)]^{2}}
{[s-2\omega_{0}-\Delta(s)]^{2}+\Delta^{2}(s)-\Delta^{\prime}(s)(s-2\omega_{0})^{2}},\\
Y&(s)=\dfrac{\Delta^{2}(s)}
{2\Delta(s)(\Delta(s)-s)+\Delta^{\prime}(s)(s^{2}-4\omega_{0}^{2})}.
\end{align}
\end{subequations}
Both the dissipationless oscillations arose from the localized modes (localized bound states) and the non-exponential decays in 
Eq.~(\ref{exactU}) are fully determined by the spectral density.

On the other hand, $\{\boldsymbol{F}_{k}(t,t_{0})\}$ in Eq.~(\ref{EOM}) is the noise source which characterizes the noise force, 
the right-hand side of Eq.~(\ref{EOMa}) associated with the initial operators $\{b_{k}(t_{0}),b^{\dagger}_{k}(t_{0})\}$ 
of the spin chain, and obeys the equation of the motion:
\begin{align}
&\dfrac{d}{dt}\boldsymbol{F}_{k}(t,t_{0})-2i\bold\omega_{0}\begin{pmatrix}
1&0\\
0&-1
\end{pmatrix}\boldsymbol{F}_{k}(t,t_{0})\notag\\
&+\int_{t_{0}}^{t}\boldsymbol{G}(t,\tau)\boldsymbol{F}_{k}(\tau,t_{0})d\tau=iV_{k}\begin{pmatrix}
e^{-2i\epsilon_{k}\tau}&e^{2i\epsilon_{k}\tau}\\
-e^{-2i\epsilon_{k}\tau}&-e^{2i\epsilon_{k}\tau}
\end{pmatrix}.
\end{align}
It is easy to find that its general solution is given by
\begin{align}
\boldsymbol{F}_{k}(t,t_{0})=&iV_{k}\int_{t_{0}}^{t}\boldsymbol{U}(\tau,t_{0})\begin{pmatrix}
e^{-2i\epsilon_{k}\tau}&e^{2i\epsilon_{k}\tau}\\
-e^{-2i\epsilon_{k}\tau}&-e^{2i\epsilon_{k}\tau}
\end{pmatrix}d\tau,
\label{F}
\end{align}
which generates the non-equilibrium correlation Green function $\boldsymbol{V}(t,\tau)$:
\begin{align}
&\boldsymbol{V}(t,\tau)\notag\\
&=\sum\limits_{k}\langle\boldsymbol{F}_{k}^{\dagger}(\tau,t_{0})\left(\begin{array}{c}
b_{k}^{\dagger}(t_{0})\\
b_{k}(t_{0})
\end{array}\right)\left(
b_{k}(t_{0})\ b_{k}^{\dagger}(t_{0})
\right)\boldsymbol{F}_{k}(t,t_{0})\rangle.
\end{align}
The non-equilibrium correlation Green function $\boldsymbol{V}(t,\tau)$ describes the particle-hole and particle-particle 
correlations arose from the fluctuations of the spin chain. Notice that if the flipping energy of the probing spin $\sigma_{0}$ 
equals zero ($\omega_{0}=0$), then as shown in our previous work \cite{PRB2018}, the non-equilibrium Green functions 
$\boldsymbol{U}(t,t_{0})$ and $\boldsymbol{V}(\tau,t)$ obey the following identities
\begin{subequations}
\begin{align}
&\boldsymbol{U}_{11}(t,t_{0})=\boldsymbol{U}_{22}(t,t_{0}),\ \boldsymbol{U}_{12}(t,t_{0})=\boldsymbol{U}_{21}(t,t_{0}),\notag\\
&\boldsymbol{V}_{11}(\tau,t)=\boldsymbol{V}_{22}(\tau,t)=-\boldsymbol{V}_{12}(\tau,t)=-\boldsymbol{V}_{21}(\tau,t).
\end{align}
\end{subequations}
Thus, the time-dependent
dissipation and fluctuation coefficients in the master equation Eq.~(\ref{master eq}) are reduced to
\begin{align}
\gamma(t,t_{0})=\tilde{\gamma}(t,t_{0})=-\Lambda(t,t_{0})=[\dot{\boldsymbol{U}}(t,t_{0})\boldsymbol{U}^{-1}(t,t_{0})]_{12},
\label{independent}
\end{align} 
that are solely determined by the retarded Green function $\boldsymbol{U}(t,t_{0})$. As a consequence, 
the dynamics process will be independent of the initial state of the spin chain if $\omega_{0}=0$.

\subsection{Decoherence dynamics for different phase of the spin chain}
\begin{figure}[ht]
\includegraphics[scale=0.53]{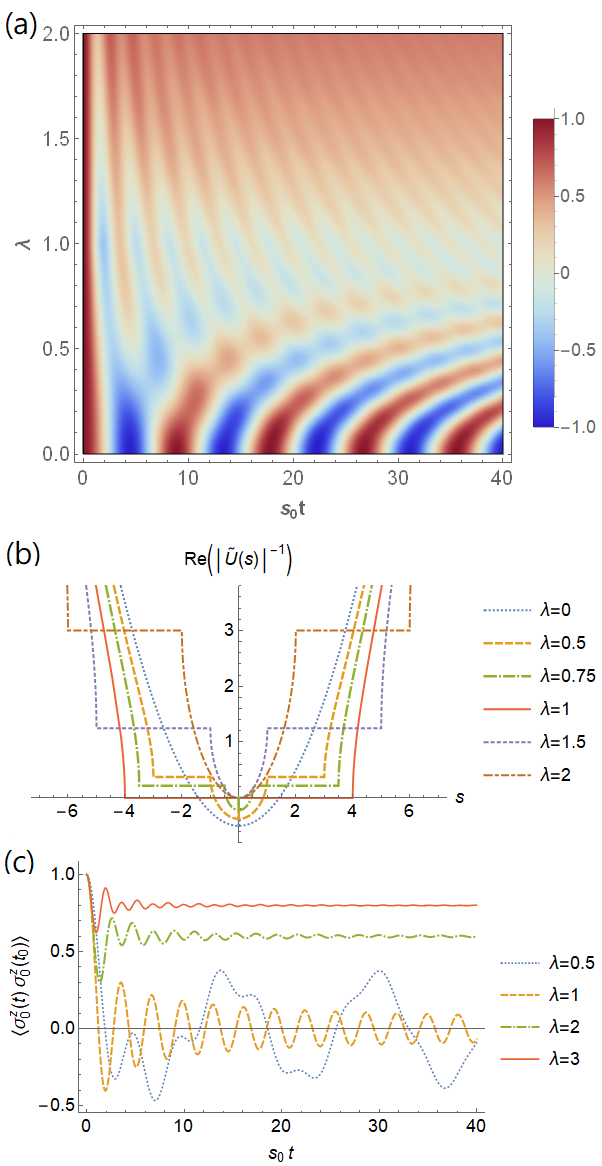}
\caption{(Colour online) (a) A contour plot of the two-time correlation $\langle\sigma_{0}^{z}(t)\sigma_{0}^{z}(t_{0})\rangle$ 
by varying the time t and $\lambda$ for $\eta=J$ and $\bold\omega_{0}=0$, where $s_{0}=2J/\hbar$. 
(b) The inverse of the determinant of $\tilde{\boldsymbol{U}}(s)$ with different $\lambda$. 
(c) The two-time correlation $\langle\sigma_{0}^{z}(t)\sigma_{0}^{z}(t_{0})\rangle$ with $\eta=J$ and $\bold\omega_{0}=0$ 
for different values of $\lambda$.} \label{fig4}
\end{figure}

Through the relation between the time-dependent dissipation $\gamma(t,t_{0})$, $\Lambda(t,t_{0})$ and the fluctuation coefficients 
$\tilde{\gamma}(t,t_{0})$ in the exact master equation and the non-equilibrium retarded and correlation Green functions,
$\boldsymbol{U}(t,t_{0})$ and $\boldsymbol{V}(t,\tau)$,  we can analytically 
solve the non-Markovian dynamics of the probing spin $\sigma_{0}$, from which the topological dynamics 
of the spin chain can be manifested in the real-time domain. As we also 
discussed earlier, the spin chain undergoes a topological phase transition from the topologically non-trivial 
phase to the topologically trivial phase when $\lambda$ changes across the critical point $\lambda_{c}=1$. 
To understand the manifestation of the topological phase transition in terms of  the real-time non-Markovian decoherence 
dynamics of the probing spin $\sigma_{0}$, we study the two-time spin correlation 
\begin{align}
\langle\sigma_{0}^{z}(t)\sigma_{0}^{z}(t_{0})\rangle
=&\langle4a^{\dagger}(t)a(t)a^{\dagger}(t_{0})a(t_{0})\rangle-\langle2a^{\dagger}(t)a(t)\rangle\notag\\
&-\langle2a^{\dagger}(t_{0})a(t_{0})\rangle+1
\end{align}
by varying the value of $\lambda$. For simplicity, we first set $\bold\omega_{0}=0$ and $\eta=J$. 
The result of the correlation $\langle\sigma_{0}^{z}(t)\sigma_{0}^{z}(t_{0})\rangle$ with different $\lambda$ 
is shown in Fig.~\ref{fig4}(a). Figure \ref{fig4}(a) shows clearly a critical transition at $\lambda=1$. 
The two-time correlation keeps oscillation between the positive and negative value in all the time for 
$\lambda<1$, while although it also oscillates for $\lambda>1$ in the beginning,
it will eventually approach to a stationary value. 

To understand the underlying mechanism of this transition, we plot the real part of the determinant of $\tilde{\boldsymbol{U}}^{-1}(s)$ 
with different $\lambda$ in Fig.~\ref{fig4}(b), which determines the localized modes of the probing spin. Notice that there 
are discontinuous parts (the flat lines) in the function due to the non-zero values of the imaginary part of 
$\tilde{\boldsymbol{U}}^{-1}(s)$ in these regions, and the imaginary part is determined by the spectral density 
$\mathcal{J}(s)$ in Eq.~(\ref{spectral}). The discontinuous parts locate exactly in the regions where the spectral 
density has non-zero value. As we have discussed earlier, the poles $\{s_{p}\}$ that make 
$|\tilde{\boldsymbol{U}}(s_{p})|^{-1}=0$ form the localized modes and contribute the dissipationless term 
in Eq.~(\ref{exactU}). The locations of these poles are the intersection points of 
$\operatorname{Re}[|\tilde{\boldsymbol{U}}(s)|^{-1}]$ and the horizontal axis with the spectral density 
$\mathcal{J}(s)=0$ (the imaginary part of $|\tilde{\boldsymbol{U}}(s)|^{-1}$ vanishes). In other words, 
the different decoherence dynamics associated with the topological phase transition is determined by 
these different dissipationless-localized modes. 

More specifically, we first consider the case of $\lambda=0$ 
that the probing spin $\sigma_{0}$ is only coupled to the left Majorana zero mode because it perfectly locates 
at the left end of the spin chain, as we have shown in Sec.~\ref{sec2}. In this case, there are three localized 
modes (one pole located at 0 and two symmetrically located at the positive and negative sides, as shown in 
Fig.~\ref{fig4}(b)). Then the energy keeps exchange between the probing spin $\sigma_{0}$ and the 
zero-energy bogoliubon of the spin chain through the left Majorana zero mode. This leads to the two-time 
correlation as a cosinusoidal oscillation for $\lambda=0$, as we can see in Fig.~\ref{fig4}(a). 

Once $\lambda>0$, 
the probing spin $\sigma_{0}$ will couple to not only the left Majorana zero mode but also others with higher 
energy modes in the spin chain, so its energy will also dissipate to the non-zero continuous modes of the spin chain. 
This leads to a non-exponential decay given by the latter term in Eq.~(\ref{exactU}). In fact, in the topologically 
non-trivial phase ($\lambda<1$), Fig.~\ref{fig4}(b) shows that there are always three localized modes. Therefore, 
after a short-time decay, the two-time correlation will reduce to a dissipationless oscillation. As $\lambda$ increasing, 
the decay term will become more and more dominant. When it reaches to the critical point $\lambda=1$, 
all the localized modes vanish (see Fig.~\ref{fig4}(b) and Fig.~\ref{fig3}). Thus, the dissipationless term vanishes 
in Eq.~(\ref{exactU}), and the spin correlation shows the maximum decoherence effect. 

On the other hand, 
in the topologically trivial phase ($\lambda>1$), Fig.~\ref{fig4}(b) shows that only one localized mode occurs 
at $s_{p}=0$. This leads to the two-time correlation eventually approach to a stationary value (no oscillation). 
Furthermore, if $\lambda$ keeps increasing, the coupling term $\eta\sigma_{0}\sigma_{1}$ between the two 
subsystems in the total Hamiltonian becomes relatively weak, which results in the two subsystems being loosely 
affected to each other. As a result, we can see that the dynamics of the two-time correlation becomes more 
and more stable as $\lambda$ gets lager and larger, as shown in Fig.~(\ref{fig4}(c)).

\begin{figure*}[ht]
\includegraphics[scale=0.43]{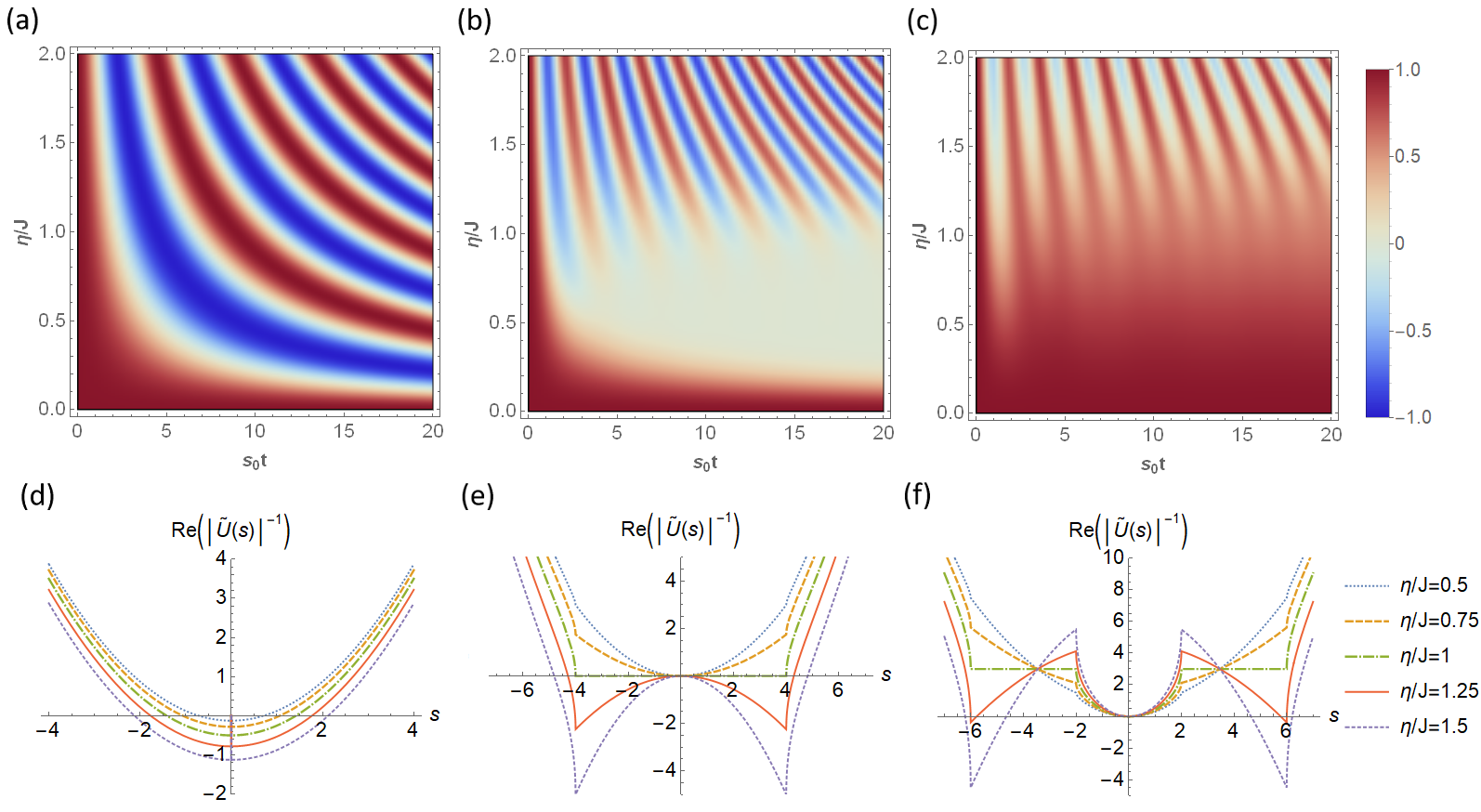}
\caption{(Colour online) The contour plot of the two-time spin-spin correlation $\langle\sigma_{0}^{z}(t)\sigma_{0}^{z}(t_{0})\rangle$ by
varying the time t and $\eta/J$ for $\bold\omega_{0}=0$ with (a)$\lambda=0$, (b)$\lambda=1$, and (c)$\lambda=2$, 
respectively, and the inverse of the determinant of $\tilde{\boldsymbol{U}}(s)$ with different $\eta$ for (d)$\lambda=0$, 
(e)$\lambda=1$, and (f)$\lambda=2$, respectively.} \label{fig5}
\end{figure*}

\subsection{Decoherence dynamics for different spin-spin chain coupling}
Notice that the coupling $\eta$ between the two subsystems can significantly affect the non-Markovian decoherence 
dynamics which is fully determined by the density of states of the spin chain and the coupling between the probing 
spin and the spin chain through the spectral density Eq.~(\ref{spectral}). In Fig.~\ref{fig5}(a), \ref{fig5}(b), and \ref{fig5}(c), 
we plot the two-time correlation $\langle\sigma_{0}^{z}(t)\sigma_{0}^{z}(t_{0})\rangle$ with different values of $\eta$ 
for the topologically non-trivial phase ($\lambda=0$), the critical point ($\lambda=1$), and the topologically trivial 
phase ($\lambda=2$), respectively. To understand these different behaviors of the correlations in different coupling 
region, we plot again the real part of the determinant of $\tilde{\boldsymbol{U}}^{-1}(s)$ with the different corresponding 
values of $\eta$ in Fig.~\ref{fig5}(d), \ref{fig5}(e), and \ref{fig5}(f). Figure \ref{fig5}(d) shows that for the topologically 
non-trivial phase ($\lambda=0$), there exist always three localized modes, which are independent of the value 
of $\eta$ (except for the trivial case $\eta=0$). The two-time correlation always shows a cosinusoidal oscillation. 
Figure \ref{fig5}(d) also shows that the change of $\eta$ will affect the locations of the localized modes, which 
determine the frequencies of the dissipationless oscillation. Because the stronger the coupling $\eta$ is, 
the easier it is to exchange energy between the two subsystems and the probing spin is affected from the 
topologically non-local state of the spin chain, the two-time correlation shows the oscillation with the higher 
frequency, as shown in Fig.~\ref{fig5}(a). 

At the critical point ($\lambda=1$), Fig.~\ref{fig5}(e) shows that 
when $\eta<J$, the real part of $|\tilde{\boldsymbol{U}}^{-1}|(s)$ vanishes only at $s=0$ where the imaginary 
part has non-zero value (the spectral density $\mathcal{J}(0)>0$, see Fig.~\ref{fig3}). Hence, there is no pole 
(localized mode) in this region, and the two-time correlation decays to zero monotonically, as a typical Markov 
process. On the other hand, we find from Fig.~\ref{fig5}(e) that there are two localized modes when $\eta>J$, 
and thus the dissipationless oscillation terms get contribution in the two-time correlation. As a result, 
the dynamics of the probing spin $\sigma_{0}$ undergoes a transition from a Markovian process in the weak 
coupling region to a non-Markovian process in the strong coupling region, as shown in Fig.~\ref{fig5}(b). 

For the topologically trivial phase ($\lambda=2$), Fig.~\ref{fig5}(f) show that there is only one localized mode 
at $s=0$ in the weak coupling region as we mentioned in Fig.~\ref{fig4}. But there are three localized modes in the 
strong coupling region. Note that the intersection points located between $2<|s|<6$ in Fig.~\ref{fig5}(f) are not 
poles because the imaginary part of $|\tilde{\boldsymbol{U}}^{-1}|(s)$ has non-zero value in this range, 
as shown in Fig.~\ref{fig3}. In conclusion, the two subsystems exchange energy in the beginning in the weak 
coupling region ($\eta<J$), and then they reach the qualitatively different steady states for the different topological 
phases of the spin chain. When the coupling $\eta$ between the two subsystems gets stronger, the probing spin 
has to take longer time to reach the steady state. While, the two subsystems always maintain energy exchange 
in the both phases of the spin chain in the strong coupling region ($\eta>J$). Thus, the topological effect of the 
spin chain to the decoherence dynamics of the probing spin becomes insignificant. 

\subsection{Decoherence dynamics for different spin-flip energy}

\begin{figure}[hbt]
\includegraphics[scale=0.30]{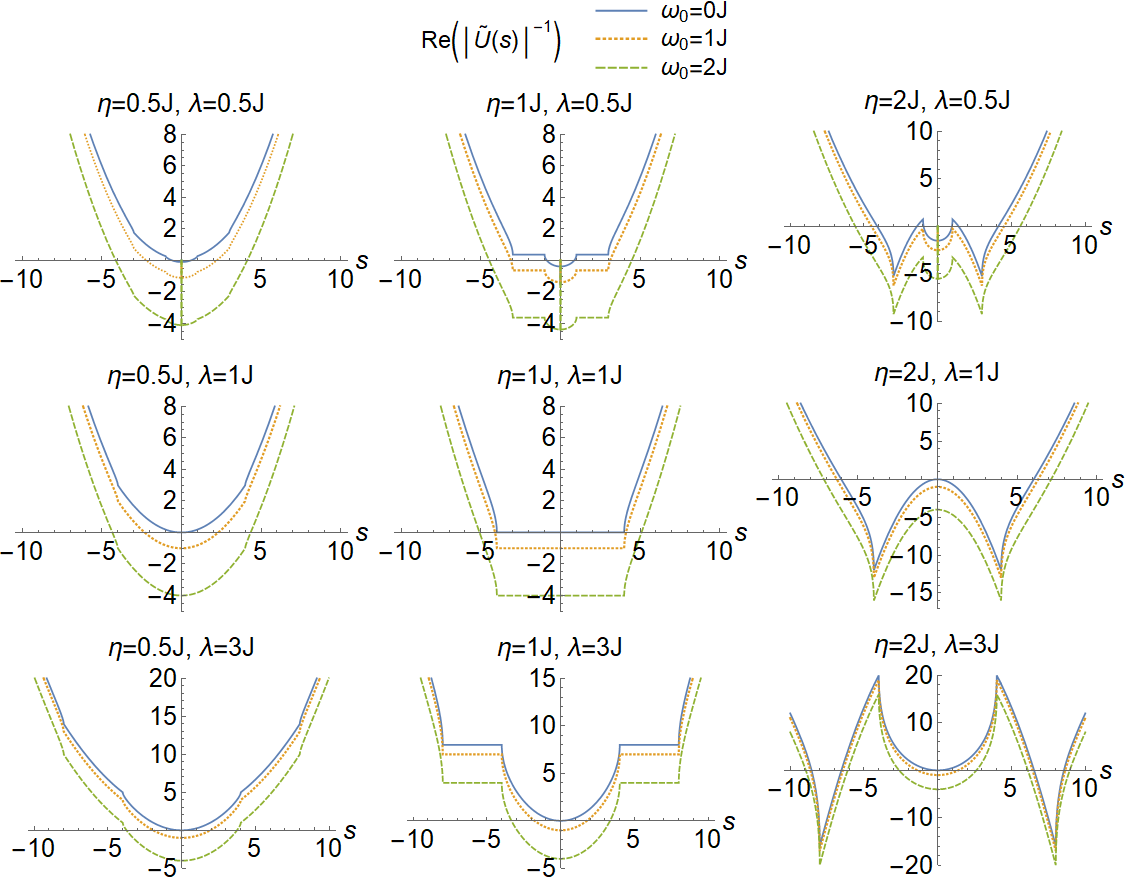}
\caption{(Colour online) The inverse of the determinant of $\tilde{\boldsymbol{U}}(s)$ with different values of 
$\omega_{0}$ in both phases and different coupling regions.} \label{fig6}
\end{figure}

\begin{table}[!htbp]
\centering
\begin{tabular}{lccc}
\hline
\hline
\quad &$\lambda<1$&$\lambda=1$&$\lambda>1$\\
\hline
$\eta\leq J, \omega_{0}=0$& 3+0& 0+0&1+0\\
\hline
$\eta\leq J, 0<\omega_{0}\leq\omega_{1}$& 3+0& $\times$&2+0\\
\hline
$\eta\leq J,\omega_{1}<\omega_{0}<\omega_{2}$& 1+0& 0+0&0+0\\
\hline
$\eta\leq J,\omega_{2}\leq\omega_{0}$&1+2 &0+2&0+2\\
\hline
$\eta>J,\omega_{0}=0$&3+2 &0+2&1+2\\
\hline
$\eta>J,0<\omega_{0}\leq\omega_{1}$&3+2 &$\times$&2+2\\
\hline
$\eta>J,\omega_{1}<\omega_{0}$&1+2 &0+2&0+2\\
\hline\hline
\end{tabular}
\caption{The number of localized modes with different values of $\omega_{0}$, $\eta$, and $\lambda$.}\label{Table1}
\end{table}

\begin{figure}[ht]
\includegraphics[scale=0.30]{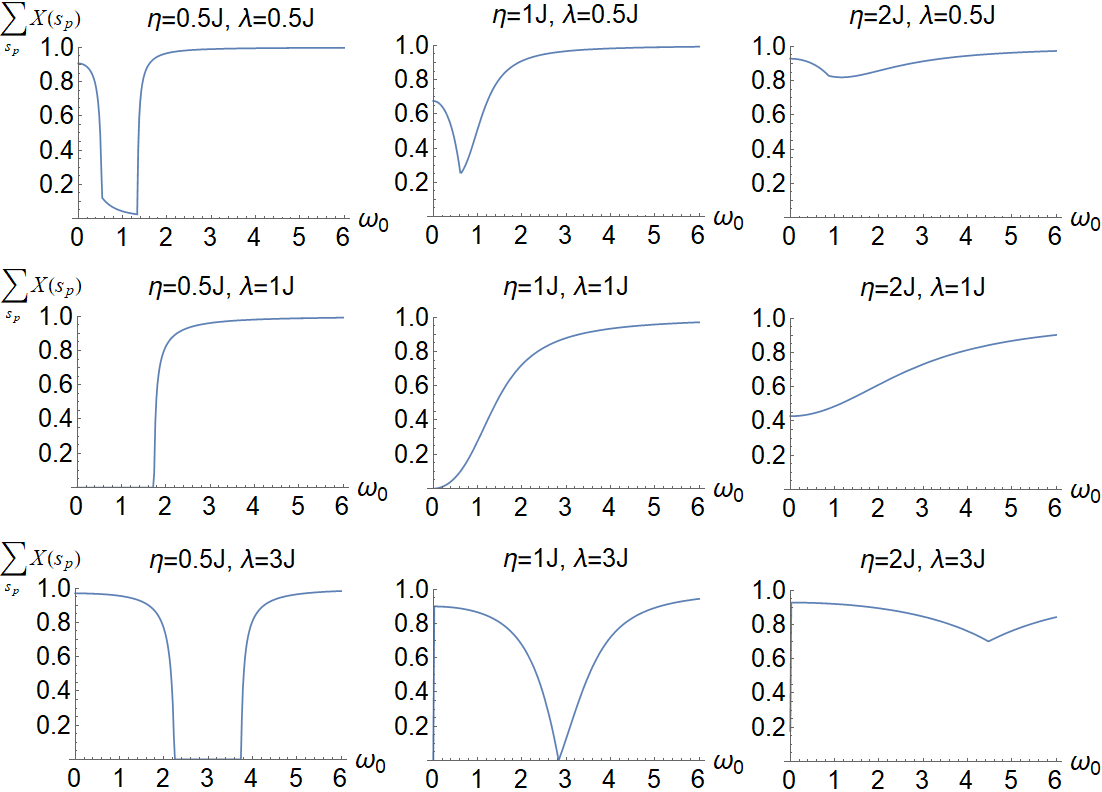}
\caption{(Colour online) The summation of the amplitudes of all localized modes versus $\omega_{0}$ in each 
phases and coupling regions.} \label{fig7}
\end{figure}

In the previous discussion, we only discuss about the case that the probing spin $\sigma_{0}$ has zero flipping 
energy ($\omega_{0}=0$), in which the time-dependent coefficients in the master equation are independent of 
the correlation Green function $\boldsymbol{V}(t,t)$, as shown in Eq.~(34). In other words, for $\omega_{0}=0$, 
the decoherence dynamics is independent of the environmental noise which is associated with the initial state of 
the spin chain. To have a further understanding of the effect of the spin-flip energy $\omega_{0}$ to the decoherence dynamics of the  
spin $\sigma_{0}$ and the consequence of the initial dependence of the spin chain, we first plot the real part of the 
determinant of $\tilde{\boldsymbol{U}}^{-1}(s)$ again for topologically non-trivial phase, topologically trivial phase, 
and critical point in different coupling regions with different values of $\omega_{0}$ in Fig.~\ref{fig6}. Figure \ref{fig6} 
shows that the increase of $\omega_{0}$ always makes the locations of the localized modes away from zero, which 
may affect the number of localized modes, as listed in Table~\ref{Table1}. For the localized modes numbers $a+b$ listed 
in Table~\ref{Table1}, $a$ is the number of the localized modes located between $-2|1-\lambda|<s<2|1-\lambda|$ and 
$b$ is that in $s<-2(1+\lambda)$ or $s>2(1+\lambda)$. Notice that these numbers will change when $\omega_{0}$ crosses $\omega_{1}$ or $\omega_{2}$, where
\begin{subequations}
\label{omega}
\begin{align}
\omega_{1}&=\left\{\begin{array}{l}
\sqrt{(1-\lambda)(1-\lambda-\eta^{2}(1-3\lambda)/2J^{2})}\quad\lambda<1\\ ~ \\
\sqrt{(\lambda-1)(\lambda-1+2\eta^{2}/J^{2})}\hfill\lambda\geq1
\end{array} \right.\\
\omega_{2}&=\left\{\begin{array}{l}
\sqrt{(1+\lambda)(1+\lambda-\eta^{2}(1+3\lambda)/2J^{2})}\quad\lambda<1\\ ~\\
\sqrt{(1+\lambda)(1+\lambda-2\eta^{2}/J^{2})}\hfill\lambda\geq1
\end{array} \right. .
\end{align}
\end{subequations}
For special case $\lambda=1$, we have $\omega_{1}=0$, and for the case $\eta=J$, we have $\omega_{1}=\omega_{2}$. 
Moreover, the amplitudes of the localized modes located in these two regions have contrast behavior as $\omega_{0}$ 
increasing. The amplitudes of the localized modes decrease in the region $-2|1-\lambda|<s<2|1-\lambda|$, while they 
increase in the region $s<-2(1+\lambda)$ or $s>2(1+\lambda)$. This can be seen clearly from Fig.~\ref{fig7} where 
the summation of the amplitudes of all localized modes versus $\omega_{0}$ is plotted.

\begin{figure*}[htb]
\includegraphics[scale=0.48]{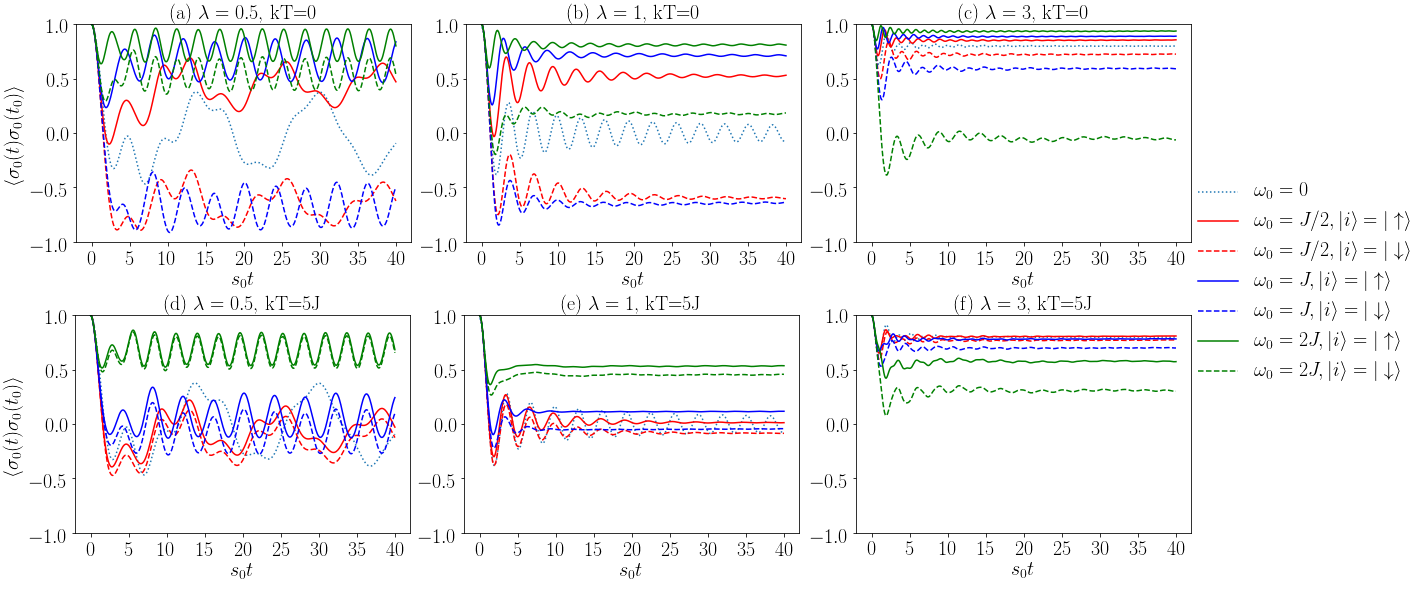}
\caption{(Colour online) The two-time correlation $\langle\sigma_{0}^{z}(t)\sigma_{0}^{z}(t_{0})\rangle$ with different 
initial states of $\sigma_{0}$ and different values of $\omega_{0}$ for (a)$\lambda=1/2$, (b)$\lambda=1$, (c)$\lambda=3$ 
at zero temperature and for (d)$\lambda=1/2$, (e)$\lambda=1$, (f)$\lambda=3$ at temperature $T=5J/k_{B}$.} \label{fig8}
\end{figure*}

Nevertheless, for $\omega_{0}\neq0$, the decoherence dynamics is no longer solely determined by the dissipation 
from $\boldsymbol{U}(t,t_{0})$, but also the fluctuations associated with the initial states of both the probing spin 
$\sigma_{0}$ and the spin chain through the correlation Green function $\boldsymbol{V}(t,t)$. 
To obtain a more comprehensive physical picture, we plot the two-time correlation $\langle\sigma_{0}(t)\sigma_{0}(t_{0})\rangle$ 
with different values of $\omega_{0}$ and different initial states of $\sigma_{0}$ for the spin chain in the topologically 
non-trivial phase ($\lambda=1/2$), the critical point ($\lambda=1$), and the topologically trivial phase ($\lambda=3$),
see  Fig.~\ref{fig8}(a), \ref{fig8}(b), and \ref{fig8}(c), respectively. The results show that if the probing spin $\sigma_{0}$ 
is initially in the high-energy (spin-down) state, its energy will be dissipated into the spin chain, and it tends to decay 
to the low-energy (spin-up) state. If the probing spin $\sigma_{0}$ is initially in the low-energy state, it exchanges little 
energy with the spin chain and most likely remains in the low-energy state. The larger the value of $\omega_{0}$ is, the 
more apparent this phenomenon can be seen in the region $\omega_{0}<\omega_{1}$, where the localized modes are all located 
between $-2|1-\lambda|<s<2|1-\lambda|$. However, for $\omega_{0}>\omega_{2}$, the amplitudes of the localized modes 
become large with increasing $\omega_{0}$, so that even if the probing spin is in the high-energy state, it becomes more hardly 
to dissipate its energy into the spin chain (see the dashed green lines in Fig.~\ref{fig8}(a) and Fig.~\ref{fig8}(b)). 

Moreover, for $\omega_{0}\neq0$, the initial state of the spin chain also affect on the decoherence dynamics of 
the probing spin $\sigma_{0}$. The spin chain is assumed initially in thermal equilibrium state. We plot again the 
two-time correlation for different phases with the spin chain at initial finite temperature $k_BT=5J$ in 
Fig.~\ref{fig8}(d), \ref{fig8}(e), and \ref{fig8}(f). The results show that for the high temperature, the two diagonal 
terms of the correlation Green function $\boldsymbol{V}_{11}(t,t)$ and $\boldsymbol{V}_{22}(t,t)$ are similar. 
As a result, the two-time correlations of the two initial states become closer to each other, as shown in 
Fig.~\ref{fig8}(d), \ref{fig8}(e), and \ref{fig8}(f). In other words, if the spin chain is initially at a relatively high 
temperature, the dependence of the non-Markovian decoherence dynamics on the initial state of $\sigma_{0}$ will diminish because the 
thermal fluctuation dominates the non-Markovian decoherence dynamics.

Putting all the above analyses together, we find that all the parameters $\lambda$ , $\eta$, and $\omega_{0}$ can 
induce different number of the localized modes with different amplitudes and therefore affect differently 
the non-Markovian decoherence dynamics associated with the topological states. 
In particular, for $\omega_{0}=0$ and in the weak coupling region, the topological phase transition can be 
significantly manifested in the dissipation dynamics of the probing spin $\sigma_{0}$. In other words, the 
topological structure of the spin chain can be observed through the non-Markovian decoherence dynamics 
of the probing spin $\sigma_{0}$. On the other hand, in the strong coupling region, the topological non-local state is more 
strongly coupled to the probing spin so that the topological effect in the non-Markovian dynamics becomes 
more significant, as shown in Fig.~\ref{fig5}. However, for $\omega_{0}\neq0$, the noise effect gets involved 
into the decoherence dynamics, which is strongly correlated with the initial state of spin $\sigma_{0}$ and 
the initial temperature of the spin chain. As a result, the manifestation of the topological structure of the spin 
chain on the non-Markovian decoherence dynamics of the probing spin $\sigma_{0}$ is merged. Hence, 
we propose the experimental probe of the topological structure of the spin chain through the decoherence 
dynamics of an external spin $\sigma_{0}$ coupling {\it weakly} to the spin chain at low temperature.

\section{The dynamics of entanglement entropy}
\label{sec5}
In this section, we study the dynamics of the quantum entanglement. In the static case, the behavior of the 
entanglement has a universal character that the entanglement of the system state would be enhanced near a 
quantum phase transition and reach the maximum at the critical point. Therefore, it can be used as an 
estimator of quantum correlations \cite{EntangleManyBody} and as a detector to classify quantum phase 
transitions \cite{quench1,Nature2002,Cardy2004,DynamicEntangle}. It is also 
interesting to see how entanglement developed in time when the system is far away from the equilibrium 
state or ground state. Therefore, we would like to further investigate the relation between the entanglement 
and quantum phase transitions in the non-equilibrium region.

The entanglement between the probing spin $\sigma_{0}$ and the spin chain can be characterized by 
the von Neumann entropy $S_{A}(t)=-\Tr [\rho_{A}(t)\ln\rho_{A}(t)]$. The reduced density matrix $\rho_{A}(t)$ 
of the probing spin $\sigma_{0}$ that obeys the master equation Eq.~(\ref{master eq}) can be expressed as
\begin{subequations}
\begin{align}
(\rho_{A})_{11}(t)=&\boldsymbol{V}_{22}(t,t)+\boldsymbol{U}_{12}(t,t_{0})\boldsymbol{U}_{21}(t,t_{0})\langle a^{\dagger}(t_{0})a(t_{0})\rangle\notag\\
&+\boldsymbol{U}_{11}(t,t_{0})\boldsymbol{U}_{22}(t,t_{0})\langle a(t_{0})a^{\dagger}(t_{0})\rangle\\
(\rho_{A})_{22}(t)=&\boldsymbol{V}_{11}(t,t)+\boldsymbol{U}_{11}(t,t_{0})\boldsymbol{U}_{22}(t,t_{0})\langle a^{\dagger}(t_{0})a(t_{0})\rangle\notag\\
&+\boldsymbol{U}_{12}(t,t_{0})\boldsymbol{U}_{21}(t,t_{0})\langle a(t_{0})a^{\dagger}(t_{0})\rangle\\
(\rho_{A})_{12}(t)=&(\rho_{A})^{*}_{21}(t)=\boldsymbol{U}_{11}(t,t_{0})\langle a(t_{0})\rangle\notag\\
&\hspace*{18mm}+\boldsymbol{U}_{12}(t,t_{0})\langle a^{\dagger}(t_{0})\rangle.
\end{align} 
\end{subequations}
Moreover, we find that there is a relation between the entanglement entropy $S_{A}(t)$ and the two-time 
correlation $\langle\sigma_{0}^{z}(t)\sigma_{0}^{z}(t_{0})\rangle$ if the initial state of $\sigma_{0}$ is a pure state, which is given by
\begin{align}
S_{A}(t)=-&\dfrac{1-\langle\sigma_{0}^{z}(t)\sigma_{0}^{z}(t_{0})\rangle}{2}\ln\dfrac{1-\langle\sigma_{0}^{z}(t)\sigma_{0}^{z}(t_{0})\rangle}{2}\notag\\
-&\dfrac{1+\langle\sigma_{0}^{z}(t)\sigma_{0}^{z}(t_{0})\rangle}{2}\ln\dfrac{1+\langle\sigma_{0}^{z}(t)\sigma_{0}^{z}(t_{0})\rangle}{2}.
\end{align}

\begin{figure}[tb]
\centerline{\scalebox{0.62}{\includegraphics{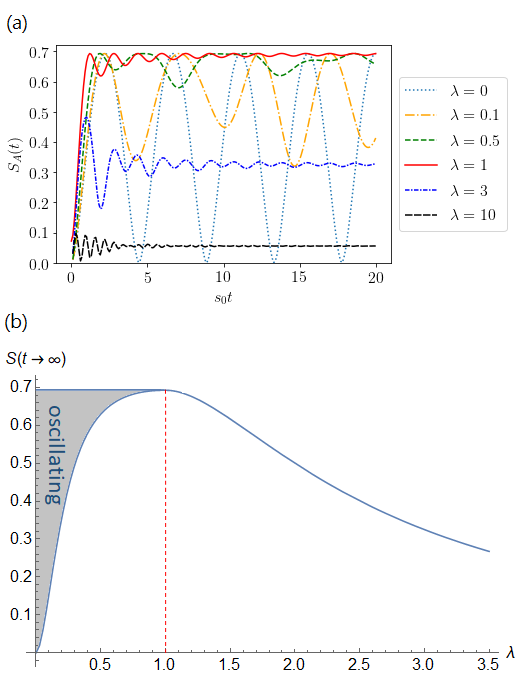}}}
\caption{(Colour online) (a) The entanglement entropy $S_{A}(t)$ for $\bold\omega_{0}=0$ and $\eta=J$ 
with the spin chain varing from the topologically non-trivial phase ($\lambda<1$) to the topologically trivial phase ($\lambda>1$). 
(b) The long-time entanglement entropy versus $\lambda$.}\label{fig9}
\end{figure}

We plot $S_{A}(t)$ for $\omega_{0}=0$ and $\eta=J$ with different values of $\lambda$ in Fig.~\ref{fig9}(a). 
In this case, we have $\boldsymbol{V}_{11}(t,t)=\boldsymbol{V}_{22}(t,t)$, then the entanglement entropy is 
independent of the initial state of the spin chain, as we mentioned in Sec.~\ref{sec3}. the probing spin $\sigma_{0}$ 
is assumed to be initially in a pure state. In the beginning, there is no entanglement between $\sigma_{0}$ and the 
spin chain, and the entropy $S_{A}(t_{0})$ equals zero. When $t>t_{0}$, the probing spin $\sigma_{0}$ begins to 
entangle with the spin chain due to the coupling between them so that the entropy $S_{A}$ increases from 0 in 
Fig.~\ref{fig9}(a). For $\lambda<1$, the entanglement entropy always keeps oscillating, which means that the 
probing spin $\sigma_{0}$ and the spin chain never reach the equilibrium state due to the existence of localized 
modes \cite{PRL2012}. The maximum value of such entanglement entropy oscillation is always $\ln 2$, 
while its lower bound rises as $\lambda$ increases. For $\lambda\geq1$, the probing spin $\sigma_{0}$ will 
reach equilibrium with the spin chain in the long-time limit, and the entanglement entropy will approach to a 
stationary value which decreases as $\lambda$ increases. Note that the entanglement entropy approaches 
to $\ln2$ at $\lambda=1$, in agreement with the expectation that there is maximum entanglement at the critical 
point of topological phase transitions. We further plot the entanglement entropy versus $\lambda$ under the 
long-time limit in Fig.~\ref{fig9}(b) to see more clearly its close relation with topological phase transition. 
Figure \ref{fig9}(b) shows a qualitative change of the entanglement entropy when the topological phase transition occurs. 
It shows that the entanglement entropy can be used to diagnose topological phase transitions in the non-equilibrium regions.

\begin{figure*}[bth]
\includegraphics[scale=0.5]{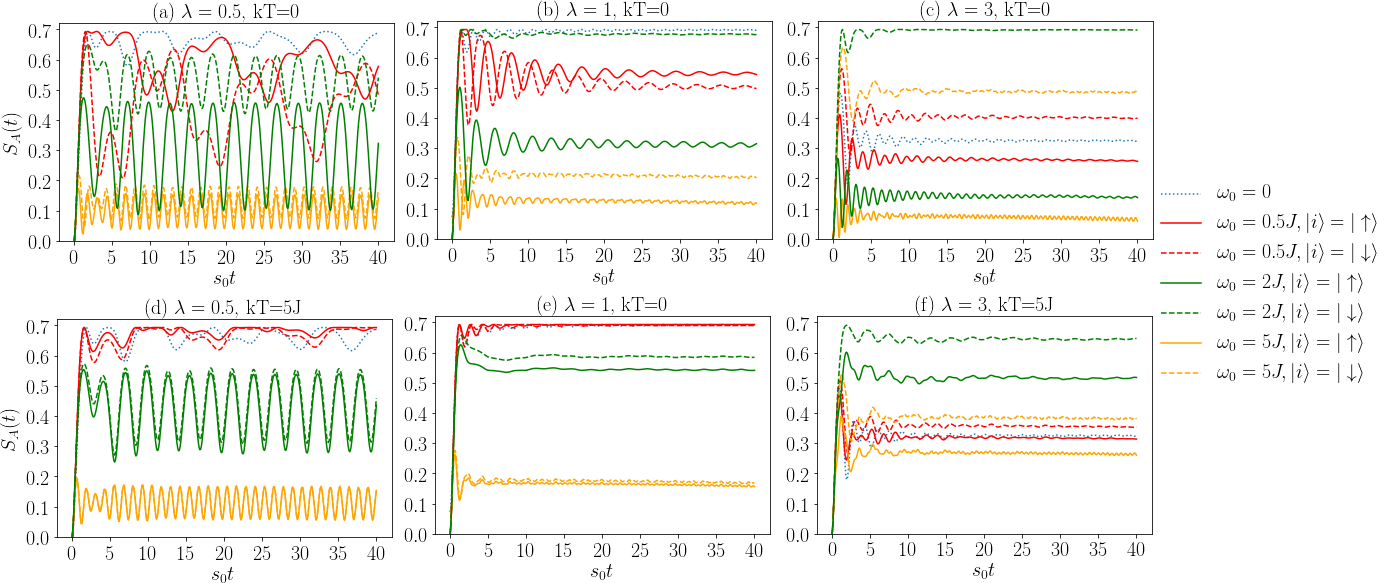}
\caption{(Colour online) The entanglement entropy $S_{A}$ with different initial states of $\sigma_{0}$ and different 
values of $\omega_{0}$ for (a)$\lambda=1/2$, (b)$\lambda=1$, (c)$\lambda=3$ in zero temperature and for 
(d)$\lambda=1/2$, (e)$\lambda=1$, (f)$\lambda=3$ in temperature $T=5J/k_{B}$.} \label{fig11}
\end{figure*}

However, for the case of $\omega_{0}\neq0$, the entanglement entropy also depends on the initial states of both
the probing spin $\sigma_{0}$ and the spin chain. The entanglement entropy $S_{A}(t)$ for the spin chain 
with the zero initial temperature in topologically non-trivial phase $\lambda<1$, critical point $\lambda=1$, 
and topologically trivial phase $\lambda>1$ is presented in Fig.~\ref{fig11}(a), \ref{fig11}(b), and \ref{fig11}(c), 
respectively. The results show that the different initial states of the probing spin $\sigma_{0}$ induce very  
different behaviors of the entanglement entropy, particularly for the high spin-flip energy $\omega_{0}$. 
This is because the probing spin $\sigma_{0}$ with the high-energy initial state is more favor to decay 
than that with the low-energy initial state. For a small value of $\omega_{0}$, the probing spin $\sigma_{0}$ 
initially in the low-energy state tends to remain in the pure initial state, while $\sigma_{0}$ which is initially 
in the high-energy state tends to decay to the mixed state. As a result, we can see in Fig.~\ref{fig11} that for 
$\omega_{0}=0.5J$ (red lines) and $\omega_{0}=2J$ (green lines), the probing spin $\sigma_{0}$ with the 
low-energy initial state $|\uparrow\rangle$ is less favor to be entangled with the spin chain, while the probing 
spin $\sigma_{0}$ with the high-energy initial state $|\downarrow\rangle$ is more favor to entangle with the 
spin chain. 

On the other hand, for a larger value of $\omega_{0}$, the probing spin $\sigma_{0}$ initially 
in the low-energy state still remains in its initial state, while the spin initially in the high-energy state tends to 
decay to the low-energy state. In both cases, the probing spin $\sigma_{0}$ tends to evolve toward the pure 
state so that the probing spin $\sigma_{0}$ and the spin chain are less entangled, as we can see from the
 result of $\omega_{0}=5J$ (yellow line) in Fig.~\ref{fig11}. More importantly, we can see in Fig.~\ref{fig11} 
 that for $\omega_{0}\neq0$, the critical point does not always has the maximum entanglement, i.e., 
 the enhancement of the entanglement near the critical region is suppressed in this case. We also plot the 
 entanglement entropy in different phases with the initial temperature of the spin chain $k_{B}T=5J$ in 
 Fig.~\ref{fig11}(d), \ref{fig11}(e), and \ref{fig11}(f). The results show that the dependence of the initial 
 state of $\sigma_{0}$ also diminish in the high temperature due to the thermal fluctuation. Meanwhile, 
 the thermal fluctuation makes the entanglement entropy increase, particularly for the high-energy 
 initial state $|\downarrow\rangle$ of $\sigma_{0}$. These results for $\omega_{0}\neq0$ show that the 
 relation between the entanglement entropy and topological phase transition is less obvious due to the
  initial state dependence and the thermal effect. Therefore, we find again that to properly probe the relation 
  between the entanglement and topological phase transitions through the probing spin, 
  it is crucial to control the flip-energy of the probing spin $\sigma_{0}$ small and make
  the spin chain at low temperature.

\section{Conclusion and Perspective}
\label{sec6}
The energy eigenfunctions of the conventional transverse-field Ising model, or the equivalent Kitaev model, 
cannot be analytically solved with the free end boundary condition even though in principle it is exactly solvable. 
We introduce a modified transverse-field Ising chain with zero local transverse field at the last site of the
spin chain ($h_N=0$) such that the model becomes analytically solvable. Its spectrum as well as its ground state wavefunction distribution 
with those of the ordinary transverse-field Ising model are comparable but are also distinguishable. We show that 
different from the ground states of the original spin model which has zero energy only for $\lambda<1$ for large $N$, the 
two-fold degenerate zero energy ground states (zero modes)  always exists in the modified  transverse-field Ising
model  for all the values of $\lambda$, but the modified model still has the topological phase 
transition at $\lambda=1$, namely the zero energy ground state wavefunctions have different topological properties 
for $\lambda<1$ and $\lambda>1$, which cannot be seen obviously in its spectrum. We also prove that the phase transition is associated with the change 
of the topological winding number of the ground state wavefunctions. Moreover,  in the modified model, 
the results of the ground state wavefunction distributions indicate that the right Majorana zero mode is always located
at the right end of the spin chain because $h_N=0$. The distribution of the left Majorana zero mode is the same as that of the ordinary model for 
$\lambda<1$ so that the two Majorana zero modes are non-locally separated (topologically nontrivial). 
While for $\lambda>1$ the left Majorana zero mode in the modified model moves to the right-hand side and eventually merges 
with the right Majorana zero mode such that the zero energy Majorana modes become topologically trivial.

We then propose a scheme to measure the topological structure of the modified spin chain through 
the non-Markovian decoherence dynamics of a probing spin in the real-time domain by coupling the probing spin to 
the spin chain. We derive the exact master equation of a probing spin and analyzed in detail its non-Markovian 
decoherence dynamics by studying the two-time correlation function $\langle\sigma_{0}^{z}(t)\sigma_{0}^{z}(t_{0})\rangle$. 
We find that 
\begin{itemize}
\item[(i)] in the topologically non-trivial phase, the topological non-local property induces different localized modes in
comparison with the case in the topologically trivial phase. These localized modes qualitatively change the non-Markovian 
decoherence dynamics of the probing spin so that the topological structure of the transverse-field Ising chain is manifested. 
\item[(ii)] The coupling $\eta$ between the probing spin and the the spin chain, and the flipping energy $\omega_{0}$ 
of the probing spin also affect the non-Markovian decoherence dynamics. For strong coupling $\eta$, the non-Markovian 
oscillation is dominant in both phases so that the manifestation of the topological phase transition in the non-Markovian 
decoherence dynamics of the probing spin becomes weak. While for large spin-flip energy $\omega_{0}$, the manifestation 
of topological and non-topological phases is also suppressed due to the noise effects associated with the initial state of the probing spin and the temperature of the spin chain. 
\item[(iii)] The dynamical entanglement entropy can be expressed in the two-time correlation 
$\langle\sigma_{0}^{z}(t)\sigma_{0}^{z}(t_{0})\rangle$ so that the entanglement entropy between the probing 
spin and the spin chain can characterize the topological phase transition, which is equivalent to the description
of the decoherence dynamics of the probing spin, but the later description may be more feasible for experimental 
realization.
\end{itemize} 
As a result, the topological properties of the transverse-field Ising model can be probed through the 
non-Markovian decoherence dynamics of a probing spin weakly coupling  to it, and the dynamical  
phase transition can be also explored in terms of the dynamical entanglement entropy, as long as one
 keeps the probing spin with a small spin-flip energy and the spin chain at a low temperature. 

The results presented in this work provides indeed a general way to experimentally measure  
topological properties and dynamical phase transitions in many-body systems 
in the real-time domain,  The decoherence properties of the probing spin can manifest the topological structure and 
dynamical phase transition of a many-body system because the decoherence dynamics of the probing particle is fully determined
by the spectral density $J(\omega)= 2\pi \rho(\omega) |V(\omega)|^2$ which contains all of the information of the 
many-body spectra and many-body eigenfunction distributions through the density of state $\rho(\omega)$ and the coupling 
amplitude $V(\omega)$ between the probing spin and the many-body system. Current measurement 
of topological states are mainly carried out using surface-sensitive angle-resolved photoemission spectroscopy (ARPES) 
for directly observing surface states or the scanning tunneling microscope (STM) to visualize surface states in 
terms of the quasi-particle interference pattern in the energy domain. The decoherence dynamics of the probing 
spin can be experimentally measured with the time-domain single-spin Ramsey interferometry or spin echo technique. 
Furthermore, controllable coupling between the probing spin and many-body 
systems also serves as an alternative realization of the dynamical quench for the study of nonequilibrium dynamics 
of many-body systems, in particular the dynamical quantum phase transition.  Therefore, one can measure  
topological properties and dynamical phase transitions in many-body systems 
in the real-time domain, in terms of single probing particle measurement, which should be more flexible 
in comparison with real-time many-body measurements.

\section*{Acknowledgement}
We acknowledges the support from the Ministry of Science and Technology of Taiwan under Contract No. MOST-108-2112-M-006-009-MY3.

\appendix
\section{}
We rewrite the Hamiltonian of the transverse-field Ising chain with an arbitrary magnetic field at the last site ($h_{N}=h^{\prime}$) in the matrix form
\begin{align}
H_{B}=\sum\limits_{i,j=1}^{N}(c_{i}^{\dagger}A_{ij}c_{j}+c_{i}^{\dagger}B_{ij}c_{j}^{\dagger}+h.c.),
\end{align}
where $A_{ii}=-h$, $ A_{ii+1}=A_{i+1i}=B_{ii+1}=-B_{i+1i}=-J/2$ for $i<N$, $A_{NN}=h'$ and all others are zero.

If Eq.~(5) holds, then we have
\begin{align}
[b_{k},H_{B}]=\epsilon_{k}b_{k},
\end{align}
which gives
\begin{align}
\left\{\begin{array}{ll}
\epsilon_{k}u_{ki}=\sum\limits_{j=1}^{N}(u_{kj}A_{ji}-v_{kj}B_{ji})\\
\epsilon_{k}v_{ki}=\sum\limits_{j=1}^{N}(u_{kj}B_{ji}-v_{kj}A_{ji})
\end{array}\right. .
\label{A}
\end{align}
By introduce $(\Phi_{k})_{i}=u_{ki}+v_{ki}$ and $(\Psi_{k})_{i}=u_{ki}-v_{ki}$, Eq.~(\ref{A}) can be simplified as
\begin{align}
&\left\{\begin{array}{ll}
\Phi_{k}(A-B)=\epsilon_{k}\Psi_{k}\\~\\
\Psi_{k}(A+B)=\epsilon_{k}\Phi_{k}
\end{array} \right.\\~ \notag \\
\Rightarrow&\left\{\begin{array}{ll}
\Phi_{k}(A-B)(A+B)=\epsilon_{k}^{2}\Phi_{k}\\~\\
\Psi_{k}(A+B)(A-B)=\epsilon_{k}^{2}\Psi_{k}\\
\end{array}\right. ,
\end{align}
where the relevant matrices 
\begin{subequations}
\begin{align}
&(A+B)(A-B)=\notag\\
&J^{2} \left [\begin{array}{cccccc}
\lambda^{2}+1 & \lambda & 0 & \cdots & 0 & 0 \\[+2mm]
\lambda & \lambda^{2}+1 & \lambda & \cdots & 0 & 0 \\[+2mm]
0 & \lambda & \lambda^{2}+1 & \cdots & 0 & 0 \\[+2mm]
\vdots & \vdots & \vdots & \ddots & \vdots & \vdots \\[+2mm]
0 & 0 & 0 & \cdots & \lambda^{2}+1 & \lambda^{\prime} \\[+2mm]
0 & 0&0&\cdots &\lambda^{\prime}&\lambda^{\prime2}\\
\end{array}\right]
\end{align}
and
\begin{align}
&(A-B)(A+B)=\notag\\
&J^{2} \left [\begin{array}{cccccc}
\lambda^{2} & \lambda & 0 & \cdots & 0 & 0 \\[+2mm]
\lambda & \lambda^{2}+1 & \lambda & \cdots & 0 & 0 \\[+2mm]
0 & \lambda & \lambda^{2}+1 & \cdots & 0 & 0 \\[+2mm]
\vdots & \vdots & \vdots & \ddots & \vdots & \vdots \\[+2mm]
0 & 0 & 0 & \cdots & \lambda^{2}+1 & \lambda \\[+2mm]
0 & 0&0&\cdots &\lambda&\lambda^{\prime2}+1\\
\end{array}\right].
\end{align}
\end{subequations}
These matrices have the eigenenergies
\begin{align}
\epsilon^{2}_{q}=J^{2}(1+\lambda^{2}-2\lambda\cos q),  \label{eem}
\end{align}
where all the normal modes $q$ are determined by the following transcendental equation,
\begin{align}
-\lambda^{\prime}\sin[q(N-1)]=(1+\lambda^{2}-2\lambda\cos q-\lambda^{\prime2})\sin (Nq).
\end{align}

Specifically, we have the following solutions:

\noindent (i). For the case $\lambda^{\prime}=\lambda$, Eq.~(A8) can be reduced to 
\begin{align}
\sin(qN)=\lambda\sin[q(N+1)],
\end{align}
which determines all the normal modes $q$. For $\lambda \geq 1$, there are $N$ real roots, exhausting the normal modes.
For $\lambda <1$, there are $N-1$ real roots and  one imaginary root, as shown in Ref. \cite{Lieb61,1970}. 

\noindent (ii). In our case $\lambda^{\prime}=0$ (the modified model), Eq.~(A8) is simply reduced to 
\begin{align}
(1+\lambda^{2}-2\lambda\cos q)\sin(qN)=0 ,
\end{align}
which gives a set of $N-1$ solutions for $\sin(qN)=0$, i.e.
\begin{align}
q_k=\dfrac{k\pi}{N},\quad k=1,2,\cdots,N-1
\end{align}
and a special solution corresponding to the zero eigenvalue determined by 
\begin{align}
(1+\lambda^{2}-2\lambda\cos q_0)=0
\end{align}
for arbitrary $\lambda$ value.

\noindent (iii). Note that this solution is also different from the transverse-field Ising model with period boundary condition.
For the period boundary condition, we have the additional matrix element $A_{1N}=A_{N1}=B_{1N}=-B_{N1}=-J/2$.
Then, the eigenenergies is still given by Eq.~(\ref{eem}) but the normal modes are simply determined by
\begin{align}
\sin \frac{qN}{2}=0 .
\end{align} 
This gives all the N normal modes:
\begin{align}
q_k= \frac{2\pi k}{N}, \quad k=0, 1,2,\cdots,N-1.
\end{align} 
in which the zero energy mode corresponds to $k=0$ only for $\lambda=1$.  For other $\lambda \neq 1$, $\epsilon_k \neq 0$ for  $k=0$. 
It shows that the zero energy mode in three different cases are very different.

Now we focus on our modified model ($h_N=0$). For $\epsilon_{k}\neq0$, $\Psi_{k}$ and $\Phi_{k}$ can be solved respectively from the two matrices by Eq.~(A5)
\begin{subequations}
\begin{align}
(\Psi_{k})_{j}&=\alpha_{k}\sin \frac{jk\pi}{N}\\
(\Phi_{k})_{j}&=-\beta_{k}\frac{J}{\epsilon_{k}}\left\{\sin\left[\frac{(j-1)k\pi}{N}\right]+\lambda\sin \frac{jk\pi}{N}\right\},
\end{align}
\end{subequations}
and then we can obtain the wavefunctions for the non-zero-energy bogoliubons $u_{kj}=[(\Phi_{k})_{j}+(\Psi_{k})_{j}]/2$ 
and $v_{kj}=[(\Phi_{k})_{j}-(\Psi_{k})_{j}]/2$, shown as Eq.(~7). The two constants $\alpha_{k}$ and $\beta_{k}$ 
are determined by Eq.~(A4) and the commutation relation $\{b_{k},b^{\dagger}_{k}\}=1$, and the result is 
$\alpha_{k}=\beta_{k}=2{\cal N}_{k}$, where ${\cal N}_{k}$ is shown in Eq.~(9a). 

For $\epsilon=0$, the eigenvectors of the two matrices are
\begin{subequations}
\begin{align}
(\Psi_{k_0})_{j}&=\alpha_{0}\delta_{j,N}\\
(\Phi_{k_0})_{j}&=\beta_{0}(-\lambda)^{j-1}.
\end{align}
\end{subequations}
The two constants $\alpha_{0}$ and $\beta_{0}$ is determined by the commutation relations 
$\{b_{0},b^{\dagger}_{0}\}=1$ and $\{b_{0},b_{0}\}=\{b^{\dagger}_{0},b^{\dagger}_{0}\}=0$ which yield 
\begin{align}
\left\{\begin{array}{l}
\dfrac{1}{2}\sum\limits_{j=1}^{N}[(\Phi_{k_0})_{j}^{2}+(\Psi_{k_0})_{j}^{2}]=1\\
\dfrac{1}{2}\sum\limits_{j=1}^{N}[(\Phi_{k_0})_{j}^{2}-(\Psi_{k_0})_{j}^{2}]=0
\end{array}\right. ,
\end{align}
and the results are $\alpha_{0}=1$ and $\beta_{0}=2{\cal N}_{k_0}$, where ${\cal N}_{k_0}$ is shown in Eq.~(9b). 
Thus it is easy to obtain the wavefunction for the zero-energy bogoliubon $u_{k_0j}$ and $v_{k_0j}$, shown as Eq.~(8).

\section{}
We begin with Eq.~(\ref{rho0}) to derive the master equation. After integrating over all the degrees of freedom 
of the spin chain by path integral approach, we obtain the exact form of the propagating function
\begin{align}
\mathcal{K}(\xi_{f}^{*},\xi_{f}^{\prime},t|\xi_{0},&\xi_{0}^{\prime*},t_{0})\notag\\
=\mathcal{N}(t)\exp\bigg[&
\left( \begin{array}{cc}
\xi^{*}_{f} & \xi^{\prime}_{f}
\end{array}\right)\boldsymbol{J}_{1}(t,t_{0})
\left( \begin{array}{c}
\xi_{0}\\
\xi^{\prime*}_{0}
\end{array}\right)\notag\\
+&\left( \begin{array}{cc}
\xi^{*}_{f} & \xi^{\prime}_{f}
\end{array}\right)\boldsymbol{J}_{2}(t,t_{0})
\left( \begin{array}{c}
\xi^{\prime}_{f}\\
\xi^{*}_{f}
\end{array}\right)\notag\\
+&\left( \begin{array}{cc}
\xi^{\prime*}_{0} & \xi_{0}
\end{array}\right)\boldsymbol{J}_{3}(t,t_{0})
\left( \begin{array}{c}
\xi_{0}\\
\xi^{\prime*}_{0}
\end{array}\right)\notag\\
+&\left( \begin{array}{cc}
\xi^{\prime*}_{0} & \xi_{0}
\end{array}\right)\boldsymbol{J}_{1}^{\dagger}(t,t_{0})
\left( \begin{array}{c}
\xi^{\prime}_{f}\\
\xi^{\prime*}_{f}
\end{array}\right)
\bigg],
\label{propagator}
\end{align}
where $\mathcal{N}(t)$ is the normalization constant and $\boldsymbol{J}_{1}(t,t_{0})$, 
$\boldsymbol{J}_{2}(t,t_{0})$, and $\boldsymbol{J}_{3}(t,t_{0})$ are functions of the $\boldsymbol{U}(t,t_{0})$
and $\boldsymbol{V}(t,t)$, and their exact formulas are given in Ref. \cite{PRB2018}].

After substituting Eq.~(\ref{propagator}) into Eq.~(\ref{rho0}) and taking the time derivative on both sides, we have
\begin{widetext}
\begin{align}
\langle\xi_{f}\vert\dot{\rho}_{A}(t)\vert\xi_{f}^{\prime}\rangle
=\mathcal{N}(t)\int& d\mu(\xi_{0}) d\mu(\xi_{0}^{\prime})
\langle\xi_{0}\vert\rho_{A}\vert\xi_{0}^{\prime}\rangle
\mathcal{K}(\xi_{f}^{*},\xi_{f}^{\prime},t|\xi_{0},\xi_{0}^{\prime*},t_{0})\bigg[\dfrac{\dot{\mathcal{N}(t)}}{\mathcal{N}(t)}+
\left( \begin{array}{cc}
\xi^{*}_{f} & \xi^{\prime}_{f}
\end{array}\right)\boldsymbol{\dot{J}}_{1}(t,t_{0})
\left( \begin{array}{c}
\xi_{0}\\
\xi^{\prime*}_{0}
\end{array}\right)\notag\\
+&\left( \begin{array}{cc}
\xi^{*}_{f} & \xi^{\prime}_{f}
\end{array}\right)\boldsymbol{\dot{J}}_{2}(t,t_{0})
\left( \begin{array}{c}
\xi^{\prime}_{f}\\
\xi^{*}_{f}
\end{array}\right)
+\left( \begin{array}{cc}
\xi^{\prime*}_{0} & \xi_{0}
\end{array}\right)\boldsymbol{\dot{J}}_{3}(t,t_{0})
\left( \begin{array}{c}
\xi_{0}\\
\xi^{\prime*}_{0}
\end{array}\right)
+\left( \begin{array}{cc}
\xi^{\prime*}_{0} & \xi_{0}
\end{array}\right)\boldsymbol{\dot{J}}_{1}^{\dagger}(t,t_{0})
\left( \begin{array}{c}
\xi^{\prime}_{f}\\
\xi^{\prime*}_{f}
\end{array}\right)
\bigg].
\label{dotrho}
\end{align}
\end{widetext}
The propagator $\mathcal{K}(\xi_{f}^{*},\xi_{f}^{\prime},t|\xi_{0},\xi_{0}^{\prime*},t_{0})$ acting on the Grassmann 
numbers $\xi^{\prime*}_{0}$, $\xi_{0}$ of the initial state can be transferred into functions which only depend on the 
Grassmann numbers $\xi^{\prime}_{f}$, $\xi^{*}_{f}$ of the state at time t. Then Eq.~(\ref{dotrho}) becomes
\begin{align}
\langle\xi_{f}\vert\dot{\rho}_{A}(t)\vert\xi_{f}^{\prime}\rangle
&=\langle\xi_{f}\vert\rho_{A}(t)\vert\xi_{f}^{\prime}\rangle\bigg[\dfrac{\dot{\mathcal{N}}(t)}{\mathcal{N}(t)}+A(t)
+B(t)\xi^{*}_{f}\xi^{\prime}_{f}\notag\\
&+C(t)\xi^{*}_{f}\dfrac{\partial}{\partial\xi^{\prime}_{f}}
+D(t)\xi^{*}_{f}\dfrac{\partial}{\partial\xi^{*}_{f}}
+E(t)\xi^{\prime}_{f}\dfrac{\partial}{\partial\xi^{\prime}_{f}}\notag\\
&+F(t)\xi^{\prime}_{f}\dfrac{\partial}{\partial\xi^{*}_{f}}
+G(t)\dfrac{\partial}{\partial\xi_{f}^{*}}\dfrac{\partial}{\partial\xi_{f}^{\prime}}\bigg],
\label{dotrho1}
\end{align}
where
\begin{subequations}
\label{ABC}
\begin{align}
A&=\dfrac{\dot{\mu}_{3}(\nu_{2})^{2}}{|\mu_{1}|^{2}-|\nu_{1}|^{2}}\\
B&=\dfrac{2\nu_{2}(-\dot{\mu}_{1}\mu_{1}^{*}+\dot{\nu}_{1}\nu_{1}^{*})
-\dot{\nu}_{3}(\nu_{2})^{2}}
{|\mu_{1}|^{2}-|\nu_{1}|^{2}}+\dot{\nu}_{2}\\
C&=-F^{*}=\dfrac{-\dot{\nu}_{1}\mu_{1}+\dot{\mu}_{1}\nu_{1}}
{|\mu_{1}|^{2}-|\nu_{1}|^{2}}\\
D&=E=\dfrac{\dot{\mu}_{1}\mu_{1}^{*}-\dot{\nu}_{1}\nu_{1}^{*}
+\dot{\nu}_{3}\nu_{2}}
{|\mu_{1}|^{2}-|\nu_{1}|^{2}}
\\
G&=\dfrac{-\dot{\nu}_{3}}{|\mu_{1}|^{2}-|\nu_{1}|^{2}}
\end{align}
\end{subequations}
and
\begin{subequations}
\begin{align}
\nu_{i}(t,t_{0})&=[\boldsymbol J_{i}(t,t_{0})]_{11}-[\boldsymbol J_{i}^{\dagger}(t,t_{0})]_{22}\\
\mu_{i}(t,t_{0})&=[\boldsymbol J_{i}(t,t_{0})]_{12}-[\boldsymbol J_{i}(t,t_{0})]_{21}.
\end{align}
\end{subequations}
According to the three constraints: $\Tr\rho_{A}=1$, $a^{\dagger}a^{\dagger}=aa=0$, and the eigenenergies 
of A are symmetric in sign, Eq.~(\ref{ABC}) can be reduced to
\begin{align}
A=&-\dfrac{\dot{\mathcal{N}}(t)}{\mathcal{N}(t)}\\
B=&\dot{\boldsymbol{V}}_{11}(t)-[\dot{\boldsymbol{U}}(t,t_{0})\boldsymbol{U}^{-1}(t,t_{0})\boldsymbol{V}(t,t)+H.c.]_{11}\\
C=&-F^{*}=[\dot{\boldsymbol{U}}(t,t_{0})\boldsymbol{U}^{-1}(t,t_{0})]_{12}\\
D=&E=\dot{\boldsymbol{V}}_{11}(t)-[\dot{\boldsymbol{U}}(t,t_{0})\boldsymbol{U}^{-1}(t,t_{0})\boldsymbol{V}(t,t)+H.c.]_{11}\notag\\
&\hspace*{3mm}+[\boldsymbol{U}^{-1}(t,t_{0})\dot{\boldsymbol{U}}(t,t_{0})]^{\dagger}_{11}\\
G=&\dot{\boldsymbol{V}}_{11}(t)-[\dot{\boldsymbol{U}}(t,t_{0})\boldsymbol{U}^{-1}(t,t_{0})(\boldsymbol{V}(t,t)+\boldsymbol{I})+H.c.]_{11}.
\end{align}
Then using the D-algebra of the creation and annihilation operators, we obtain the exact master equation. 
\begin{align}
\dot{\rho}_{A}(t)
=-&i[\epsilon(t)a^{\dagger}a,\rho_{A}(t)]\notag\\
+&\gamma(t)[2a\rho_{A}(t)a^{\dagger}-a^{\dagger}a\rho_{A}(t)-\rho_{A}(t)a^{\dagger}a]\notag\\
+&\tilde{\gamma}(t)[a^{\dagger}\rho_{A}(t)\sigma_{0}^{-}-a\rho_{A}(t)a^{\dagger} +a^{\dagger}a\rho_{A}(t) \notag\\
&~~~~~~~~ -\rho_{A}(t)aa^{\dagger}]\notag\\
+&\Lambda(t)a^{\dagger}\rho_{A}(t)a^{\dagger}+\Lambda^{*}(t)a\rho_{A}(t)a,
\label{master0}
\end{align}
The time-dependent dissipation and fluctuation coefficients in the master equation are
\begin{subequations}
\begin{align}
\epsilon(t)=&i\Big[D(t)-\dfrac{B(t)+G(t)}{2}\Big]\notag\\
=&\dfrac{i}{2}[\dot{\boldsymbol{U}}(t,t_{0})\boldsymbol{U}^{-1}(t,t_{0})-H.c]_{11}. \\
\gamma(t)=&\dfrac{G(t)-B(t)}{2}\notag\\
=&-\dfrac{1}{2}[\dot{\boldsymbol{U}}(t,t_{0})\boldsymbol{U}^{-1}(t,t_{0})+H.c]_{11},\\[+2mm]
\tilde{\gamma}(t)=&B(t)\notag\\
=&\dot{\boldsymbol{V}}_{11}(t)-[\dot{\boldsymbol{U}}(t,t_{0})\boldsymbol{U}^{-1}(t,t_{0})\boldsymbol{V}(t,t)+H.c.]_{11},\\
\Lambda(t)=&-C(t)\notag\\
=&-[\dot{\boldsymbol{U}}(t,t_{0})\boldsymbol{U}^{-1}(t,t_{0})]_{12},
\end{align}
\end{subequations}

\end{document}